\documentclass[showpacs,nofootinbib,showkeys,eqsecnum,prd,aps,preprint]{revtex4-1}
\usepackage{amsfonts,amsmath,amssymb,bm,graphicx}

\begin{document}

\title{{\large Regularization, renormalization and consistency conditions in
QED with }$x${\large -electric potential steps}}
\author{S.~P.~Gavrilov${}^{a,c}$}
\email{gavrilovsergeyp@yahoo.com; gavrilovsp@herzen.spb.ru}
\author{D.~M.~Gitman${}^{a,b,d}$}
\email{gitman@if.usp.br}
\date{\today }

\begin{abstract}
The present article is an important addition to the nonperturbative
formulation of QED with $x$-steps presented by Gavrilov and Gitman in Phys.
Rev. D. \textbf{93}, 045002 (2016). Here we propose a new renormalization
and volume regularization{\large \ }procedures which allow one to calculate
and distinguish physical parts of different matrix elements of operators of
the current and of the energy-momentum tensor, at the same time relating the
latter quantities with characteristics of the vacuum instability.{\large \ }%
For this purpose, a modified inner product and a parameter $\tau $ of the
regularization are introduced. The latter parameter can be fixed using
physical considerations. In the Klein zone this parameter can be interpreted
as the time of the observation of the pair production effect. In the refined
formulation of QED with $x$-steps, we succeeded to consider the backreaction
problem. In the case of an uniform electric field $E$ confined between two
capacitor plates separated by a finite distance $L$, we see that the
smallness of the backreaction implies a restriction (the consistency
condition) on the product $EL$ from above.
\end{abstract}


\affiliation{${}^{a}$Department of Physics, Tomsk State University, 634050 Tomsk, Russia;\\
${}^{b}$P.N. Lebedev Physical Institute, 53 Leninskiy prospect, 119991 Moscow,
Russia;\\
${}^{c}$Department of General and Experimental Physics, Herzen State
Pedagogical University of Russia, Moyka embankment 48, 191186
St.~Petersburg, Russia\\
${}^{d}$Institute of Physics, University of S\~{a}o Paulo, CP 66318, CEP
05315-970 S\~{a}o Paulo, SP, Brazil}

\maketitle

\section{Introduction\label{S1}}

The effect of particle creation by strong electromagnetic and gravitational
fields has\emph{\ }been\emph{\ }attracting attention for a long time. The
effect has a pure quantum nature and was first considered in the framework
of the relativistic quantum mechanics with understanding that all the
questions can be answered only in the framework of quantum field theory
(QFT). QFT with external background is to a certain extent an appropriate
model for such calculations. In the framework of such a model, particle
creation is closely related to a violation of the vacuum stability with
time. Backgrounds (external fields) that violate the vacuum stability are
electric like fields that are able to produce nonzero work when interacting
with charged particles. The following statement of the problem is usually
considered. Assuming that the initial state of the quantized field of matter
is the vacuum (often called the initial vacuum, or in-vacuum), its evolution
with time under the influence of some electric like field of finite length
and duration is studied. After the external field is switched off one can
define a new vacuum (final vacuum, or out-vacuum) and calculate pairs of
particles (out-particles) appearing above this vacuum. In the general case,
the out-vacuum differs from the in-one. Depending on the structure of the
electric like fields, different approaches for calculating the particle
creation effect (Schwinger effect) were proposed and realized. Initially,
for simplification of calculations and correspondence with the idealized
problem statement described above, the effect of particle creation was
considered for uniform time-dependent external electric fields that are
switched on and off at the initial and the final time instants,
respectively.\ In what follows, we call such kind of external field as the $%
t $-electric potential step ($t$-step in what follows).{\large \ }In such an
approach, one can neglect the fact that real in- or out-particles are
spatially localized, and represent them by some plane waves. Scattering,
particle creation from the vacuum and particle annihilation by the $t$-steps
were considered in the framework of the relativistic quantum mechanics, see
Refs. \cite{Nikis70a,Nikis79,GMR85,ruffini,GelTan16}; a more complete list
of relevant publications can be found in \cite{ruffini,GelTan16}. A general
formulation, nonperturbative with respect to the time-dependent external
background of quantum electrodynamics (QED),{\large \ }was developed in
Refs. \cite{Gitman}.

It is clear that approaches elaborated for nonperturbative calculations in
QED with $t$-step\ is not applicable to the cases{\large \ }of strong
electric fields that are concentrated in restricted space areas.{\large \ }%
However, there exist many physically interesting situations where an{\large %
\ }external background formally is not switched off. As an example, we may
mention time-independent but non-uniform electric field of a constant
direction situated in a region{\large \ }$S_{\mathrm{int}}${\large \ }%
between two planes $x=x_{\mathrm{L}}$ and $x=x_{\mathrm{R}}$, it switches
off in macroscopic regions on the left of $x=x_{\mathrm{L}}$ and on the
right of{\large \ }$x=x_{\mathrm{R}}$. Namely,%
\begin{eqnarray}
\mathbf{E}\left( X\right) &=&\mathbf{E}\left( x\right) =\left( E\left(
x\right) ,0,...,0\right) ,\;E\left( x\right) =\mathrm{const}>0,\;x\in S_{%
\mathrm{int}}=\left( x_{\mathrm{L}},x_{\mathrm{R}}\right) ,  \notag \\
E\left( x\right) &=&0\,,\;x\in S_{\mathrm{L}}=\left( x_{\mathrm{FL}},x_{%
\mathrm{L}}\right] ,\ x\in S_{\mathrm{R}}=\left[ x_{\mathrm{R}},x_{\mathrm{FR%
}}\right) ,\;x_{\mathrm{FL}}<x_{\mathrm{L}}<0,\;x_{\mathrm{FR}}>x_{\mathrm{R}%
}>0\ .  \label{f.2}
\end{eqnarray}%
{\large \ }We note that the points $x_{\mathrm{FL}}$\ and $x_{\mathrm{FR}}$\
are separated from the origin by macroscopic but finite distances. Fields of
the latter form represent a kind of spatial or, as we call them
conditionally, $x$-electric potential step ($x$-step in what follows)
\footnote{%
Potentials of an external electromagnetic field $A^{\mu }\left( X\right) $
in $d=D+1$ dimensional Minkowski spacetime parametrized by coordinates $%
X=\left( X^{\mu },\ \mu =0,1,\ldots ,D\right) =\left( t,\mathbf{r}\right) ,\
X^{0}=t,\ \mathbf{r}=\left( X^{1},\ldots ,X^{D}\right) $. We assume that the
basic Dirac particle is an electron with the mass $m$ and the charge $-e$, $%
e>0$, and the positron is its antiparticle. The electric field under
consideration accelerates electrons along the $x$ axis in the negative
direction and positrons along the $x$ axis in the positive direction.},
\begin{equation}
A^{\mu }\left( X\right) =\left( A^{0}\left( x\right) ,A^{j}=0,\
j=1,2,...,D\right) ,\ x=X^{1}.  \label{f.1}
\end{equation}%
Further, for convenience, we use the following notation: $U(x)=-eA_{0}\left(
x\right) $. The magnitude of the $x$-step is%
\begin{equation}
\Delta U=U_{\mathrm{R}}-U_{\mathrm{L}}>0,\ \ U_{\mathrm{R}}=-eA_{0}(x_{%
\mathrm{FR}}),\ \ U_{\mathrm{L}}=-eA_{0}(x_{\mathrm{FL}}).  \label{f.7}
\end{equation}%
Without loss of generality, we choose a gauge $U_{\mathrm{R}}=-U_{\mathrm{L}%
} $ which emphasizes the existing symmetry between electrons and positrons.%
{\large \ }In the case when we deal with a one-particle problem and
relativistic quantum mechanics is applicable (the case of a noncritical
step, $\Delta U<2m$) such time-independent approximation of external fields
is quite common. The critical $x$-step, $\Delta U>2m$, creates particles
from the vacuum (the Klein paradox is closely related to this process \cite%
{Klein27,Sauter31a,Sauter-pot}) and then has to be considered in framework
of QED. Some heuristic calculations of the particle creation by $x$-steps in
the framework of the relativistic quantum mechanics were presented by
Nikishov in Refs. \cite{Nikis79,Nikis70b} and later developed in Refs. \cite%
{HansRavn81,Nikis04}. One should also mention Damour's work\ \cite{Damour77}%
, which contributed significantly in applying semiclassical methods for
treating strong field problems in astrophysics. Using Damour's approach,
mean number of pairs created by a strong uniform electric field confined
between two capacitor plates separated by a finite distance was calculated
in Ref.~\cite{WongW88}. A detailed historical review can be found in Refs.
\cite{DomCal99,HansRavn81}.\ At that time, however, no justification for
such calculations from the QFT point of view was known. This was because
exact solutions of relativistic equations (both the Dirac and the
Klein-Gordon equations) with an $x$-step are known in the form of stationary
plane waves with given longitudinal momenta in the regions $S_{\mathrm{L}}$%
{\large \ }and{\large \ }$S_{\mathrm{R}}$. And it is a nontrivial problem
how to use such solutions to construct initial and final vacua that are a
global states determined on a $t$-const hyperplane.

In our recent paper \cite{x-case} we have presented a technique in the
framework of standard strong-field QED that allows one to calculate effects
of the vacuum instability in the presence of $x$-steps nonperturbatively.%
{\large \ }Our approach is based on the following considerations.{\large \ }%
Physically it make sense to believe that the field of an $x$-step, given by
Eq. (\ref{f.2}), should be considered as a part of a time-dependent
inhomogeneous electric field{\large \ }$\mathbf{E}_{\mathrm{pristine}}\left(
X\right) ${\large \ }directed along the $x$-direction,{\large \ }which was
switched on very fast before the time instant $t_{\mathrm{in}}$, by this
time it had time to spread to the whole area $S_{\mathrm{int}}$ and
disappear in the macroscopic regions $S_{\mathrm{L}}$\ and $S_{\mathrm{R}}$.
Then it was switched off very fast just after the time instant{\large \ }$t_{%
\mathrm{out}}=t_{\mathrm{in}}+T${\large . }The field $\mathbf{E}_{\mathrm{%
pristine}}\left( X\right) =\mathbf{E}\left( x\right) $ in the regions $S_{%
\mathrm{L}}$, $S_{\mathrm{int}}$ and $S_{\mathrm{R}}$ from\ $t_{\mathrm{in}}$%
{\large \ }till{\large \ \ }$t_{\mathrm{out}}$ , that is, it acts as a
constant field concentrated in a restricted space area $S_{\mathrm{int}}$%
{\large \ }(we assume standard volume regularization with respect of
hypersurface orthogonal to the $x$-direction ){\large \ }during the
sufficiently large (macroscopic)\ period of time $T$,
\begin{equation}
T\gg \left( eE\right) ^{-1/2}\max \left\{ 1,m^{2}/eE\right\} ,  \label{m27}
\end{equation}%
where $E$ is the average value of the constant field. By analogy with how
this is done in the time-independent potential scattering due to noncritical
steps, it is assumed that there exist time-independent observables in the
presence of critical $x$-steps. For example, it seems quite natural that the
pair-production rate and the flux of created particles are constant during
the time\ $T$.\ It means that a leading contribution to the number density
of created electron-positron pairs is assumed to be proportional to the
large dimensionless parameter $\sqrt{eE}T$\ and is independent from fast
switching-on and -off if this parameter satisfies Eq. (\ref{m27}).\emph{\ }%
This assumption seems quite natural as much of this behavior is observed for
results obtained in exactly solvable cases with $t$-steps \cite%
{GavG96a,AdoGavGit17,AdoGavGit18,AdoFerGavGit18} and numerical calculations;
see, e.g. \cite{KluMotEis98}\footnote{%
Note that the pair-production rate per unit volume due to homogeneous fields
($x_{\mathrm{L}}\rightarrow -\infty $, $x_{\mathrm{R}}\rightarrow \infty $)
of given average intensity\ is equal to or higher than that for the case of
a finite width $x_{\mathrm{R}}-x_{\mathrm{L}}$; see Ref. \cite{L-field}.}%
{\large . }It is clear that the process of pair creation is transient.%
{\large \ }Nevertheless, the condition of the smallness of backreaction
shows there is a window in the parameter range of $E$\ and $T$\ where the
constant field approximation is consistent{\large \ }\cite{GavG08}.

It is well known that in general the vacuum instability is manifest in the
fact that the initial (true) vacuum, $\left\vert 0,\mathrm{true\;in}%
\right\rangle $, in which the system was before turning on the electric
field, differs significantly from the final (true) vacuum, $\left\vert 0,%
\mathrm{true\;out}\right\rangle $, in which the system appears after turning
off this field. For example, this difference can be estimated considering
vacuum mean values of the longitudinal current operator \emph{\ }$%
J^{1}\left( X\right) $ in the regions $S_{\mathrm{L}}$\ and/or $S_{\mathrm{R}%
}$\emph{\ }at any time instant\emph{\ }$t_{0}$\emph{\ }from\emph{\ }\ $t_{%
\mathrm{in}}${\large \ }till{\large \ \ }$t_{\mathrm{out}}$\ . One can
calculate the density of the longitudinal current corresponding to final
particles as follows:%
\begin{eqnarray}
\bar{J}_{\mathrm{cr}}^{1} &=&\left. J_{\mathrm{cr}}^{1}\left( X\right)
\right\vert _{t=t_{0,}x\in S_{\mathrm{L/R}}},  \notag \\
J_{\mathrm{cr}}^{1}\left( X\right) &=&\left\langle 0,\mathrm{true\;in}%
\left\vert J^{1}\left( X\right) \right\vert 0,\mathrm{true\;in}\right\rangle
-\left\langle 0,\mathrm{true\;out}\left\vert J^{1}\left( X\right)
\right\vert 0,\mathrm{true\;out}\right\rangle .  \label{i1}
\end{eqnarray}%
However, both fast switching-on and -off produce electron--positron pairs
with quantum numbers of the tiny range of kinetic energy and one can neglect
these contributions to any total characteristics of particle creation, which
are determined by the sum over all kinetic energies, if the time{\large \ }$%
T $\ satisfies Eq. (\ref{m27}) \cite%
{GavG96a,AdoGavGit17,AdoGavGit18,AdoFerGavGit18}. These total
characteristics are,{\large \ }for example, the vacuum-to-vacuum transition
probability, pair-production rate, and fluxes of charge and energy of
created particles. The mean values given by Eq. (\ref{i1}) depend rather
weakly on\ switching-on and -off effects. Neglecting effects of fast
switching-on and -off for the total characteristics, one can approximate
these quantities{\large \ }using in calculations instead of the true vacua $%
\left\vert 0,\mathrm{true\;in}\right\rangle $\ and $\left\vert 0,\mathrm{%
true\;out}\right\rangle $ some time-independent states $\left\vert 0,\mathrm{%
in}\right\rangle $\ and $\left\vert 0,\mathrm{out}\right\rangle $,
respectively, closest to them in a certain sense. We choose these states as
states with a minimum kinetic energy with respect to any\emph{\ }uncharged
quasistationary state (kinetic energies of the states $\left\vert 0,\mathrm{%
in}\right\rangle $\ and $\left\vert 0,\mathrm{out}\right\rangle $ are the
same; see Ref. \cite{x-case} for details) and the mean values%
\begin{equation}
\left\langle J^{1}\right\rangle _{\mathrm{in}}=\left. \left\langle 0,\mathrm{%
in}\left\vert J^{1}\left( X\right) \right\vert 0,\mathrm{in}\right\rangle
\right\vert _{t=t_{0,}x\in S_{\mathrm{L/R}}},\;\left\langle
J^{1}\right\rangle _{\mathrm{out}}=\left. \left\langle 0,\mathrm{out}%
\left\vert J^{1}\left( X\right) \right\vert 0,\mathrm{out}\right\rangle
\right\vert _{t=t_{0,}x\in S_{\mathrm{L/R}}}  \label{i2}
\end{equation}%
represent the leading contribution to the means of Eq. (\ref{i1}). The
quantities (\ref{i2}) are independent from fast switching-on and -off. They
are proportional to the number density of partial vacuum states excited due
to an electric field, that is, they are proportional to the number density
created electron-positron pairs, and therefore they are proportional to the
large parameter $\sqrt{eE}T$. We can estimate the accuracy of the
approximation in use as\footnote{%
It would be desirable also to provide detailed numerical evaluations for
realistic space-time-dependent configurations of external fields (going
beyond a locally constant field approximations) to test the validity of the
developed approach on a quantitative level.}%
\begin{equation}
\left. \left\langle 0,\mathrm{true\;in/out}\left\vert J^{1}\left( X\right)
\right\vert 0,\mathrm{true\;in/out}\right\rangle \right\vert _{t=t_{0,}x\in
S_{\mathrm{L/R}}}=\left\langle J^{1}\right\rangle _{\mathrm{in/out}}\left[
1+O\left( \left( \sqrt{eE}T\right) ^{-1}\right) \right]  \label{i3}
\end{equation}%
\emph{\ }Note that\emph{\ }$\left\langle J^{1}\right\rangle _{\mathrm{in}}$%
\emph{\ }is positive, while\emph{\ }$\left\langle J^{1}\right\rangle _{%
\mathrm{out}}$\emph{\ }is negative\emph{\ }and the accuracy of the
approximation for\emph{\ }$J_{\mathrm{cr}}^{1}$\emph{\ }is the same.

In the following we call the approximating states $\left\vert 0,\mathrm{in}%
\right\rangle $\ and $\left\vert 0,\mathrm{out}\right\rangle $\ initial and
final vacua, respectively.{\large \ }In this approximation the field{\large %
\ }$E_{\mathrm{pristine}}\left( X\right) ${\large \ }acts as a constant
field of the $x$-step. Kinetic energies of the vacuum states\emph{\ }$%
\left\vert 0,\mathrm{in}\right\rangle $\emph{\ }and\emph{\ }$\left\vert 0,%
\mathrm{out}\right\rangle $\emph{\ }are equal,\ however,\emph{\ the }mean
fluxes of charge and energy for these states are quite distinct, which makes
it easy to distinguish between them.{\large \ }We consider electron and
positron excitations of these vacua as initial and final particles,
respectively.{\large \ }To define the vacua $\left\vert 0,\mathrm{in}%
\right\rangle $\ and $\left\vert 0,\mathrm{out}\right\rangle $\ and to
construct initial and final Fock spaces, it is sufficient to use the exact
solutions of the Dirac equation with a $x$-step; see Ref. \cite{x-case} for
details.{\large \ }Using the developed approach \cite{x-case} we succeeded
to calculate global effects (global quantities) of zero order with respect
to the radiative interaction, such that the vacuum to vacuum transition
amplitude $c_{v}$, the mean differential and the total numbers of created
particles, and so on for the number of the $x$-steps (step{\large \emph{\ }}%
between two capacitor plates,{\large \emph{\ }}Sauter, Klein, and
exponential steps); see \cite{L-field,GavGitSh17,GavGitSh19A,GavGitSh19D}.%
{\large \ }However, a number of important technical and principal questions
still need to be answered. Among them are questions such as:{\large \ }how
can physical information be extracted from average values of quasilocal
quantities, for example, from the matrix elements of operators of the
current and of the energy--momentum tensor (EMT) with respect to initial and
final vacua. To be able to answer to these questions in the present article
we refine and substantially supplement our constructions Ref. \cite{x-case}.

The constant electric field of a critical $x$-step produces constant fluxes
of created from the vacuum final particles during the time interval $T$%
{\large . }These particles created as electron--positron pairs and leave
field area $S_{\mathrm{int}}$, wherein electrons are emitted to the region $%
S_{\mathrm{L}}$ and positrons to the region $S_{\mathrm{R}}$. In these
regions the created particles have constant velocities in opposite
directions, moving away from the area $S_{\mathrm{int}}$. They form constant
longitudinal currents and energy fluxes in the regions $S_{\mathrm{L}}$\ and
$S_{\mathrm{R}}$ respectively. Since the time interval{\large \ }$T${\large %
\ }and the distances{\large \ }$\left\vert x_{\mathrm{FL}}-x_{\mathrm{L}%
}\right\vert $ and{\large \ }$\left\vert x_{\mathrm{R}}-x_{\mathrm{FR}%
}\right\vert ${\large \ }are macroscopic, one may believe that, measuring
characteristics of particles in the regions $S_{\mathrm{L}}$\ and $S_{%
\mathrm{R}}${\large , }we are able to evaluate the effect of pair creation
in the area $S_{\mathrm{int}}$\ for the time interval $T${\large . }To this
end,{\large \ }in order for particles created in the area $S_{\mathrm{int}}$
not to leave the regions $S_{\mathrm{L}}${\large \ }and{\large \ }$S_{%
\mathrm{R}}$ in the time $T$, we assume that{\large \ }$\left\vert x_{%
\mathrm{FL}}-x_{\mathrm{L}}\right\vert ,\left\vert x_{\mathrm{R}}-x_{\mathrm{%
FR}}\right\vert >T$, i.e., the regions{\large \ }$\left( -\infty ,x_{\mathrm{%
FL}}\right] $ and $\left[ x_{\mathrm{FR}},+\infty \right) ${\large \ }are
not causally related to processes in the area $S_{\mathrm{int}}$. In this
article we are interested in such quasilocal quantities as{\large \ } the
vector of electric current density, $J_{\mathrm{cr}}^{\mu }(x)$, and the
energy--momentum tensor of created particles, $T_{\mathrm{cr}}^{\mu \nu }(x)$%
, measured in the macroscopic regions $S_{\mathrm{L}}$\ and $S_{\mathrm{R}}$%
{\large . } In the general case these quantities depend on the structure of
the electric field, and, in particular, they depend on the coordinate $x$
due to the $x$-dependence of the external field.{\large \ }Our first task is
to demonstrate that in the leading term approximation these quantities can
be obtained using the vacuum matrix elements of the operators of the current
$J^{\mu }(x)$ and of the EMT $T_{\mu \nu }(x)$ with respect to the initial
and final vacua defined{\large , }according to the general formulation of
strong-field QED \cite{Gitman} (see also Refs. \cite{GMR85,PluMG86}),%
\begin{align}
& \left\langle J^{\mu }(x)\right\rangle _{\mathrm{in/out}}=-ie\text{\textrm{%
tr}}\left[ \gamma ^{\mu }S_{\text{\textrm{in/out}}}^{c}(X,X^{\prime })\right]
|_{X=X^{\prime }}\ ,  \notag \\
& \left\langle J^{\mu }(x)\right\rangle ^{c}=-ie\text{\textrm{tr}}\left[
\gamma ^{\mu }S^{c}(X,X^{\prime })\right] |_{X=X^{\prime }}\ ;  \notag \\
& \left\langle T_{\mu \nu }(x)\right\rangle _{\mathrm{in/out}}=i\text{%
\textrm{tr}}\left[ A_{\mu \nu }S_{\text{\textrm{in/out}}}^{c}(X,X^{\prime })%
\right] |_{X=X^{\prime }}\ ,  \notag \\
& \left\langle T_{\mu \nu }(x)\right\rangle ^{c}=i\text{\textrm{tr}}\left[
A_{\mu \nu }S^{c}(X,X^{\prime })\right] |_{X=X^{\prime }}\ ,  \notag \\
& A_{\mu \nu }=\frac{1}{4}\left[ \gamma _{\mu }(P_{\nu }+P_{\nu }^{\prime
\ast })+\gamma _{\nu }\left( P_{\mu }+P_{\mu }^{\prime \ast }\right) \right]
,  \label{m5.3b}
\end{align}%
where we use a generalized causal in--out propagator $S^{c}(X,X^{\prime }),$%
{\large \ }and the so-called in--in propagator{\large \ }$S_{\text{\textrm{in%
}}}(X,X^{\prime })${\large \ }and out--out propagator{\large \ }$S_{\text{%
\textrm{out}}}(X,X^{\prime })${\large ,}
\begin{align}
& S^{c}(X,X^{\prime })=i\left\langle 0,\mathrm{out}\right\vert \hat{T}\hat{%
\Psi}(X)\hat{\Psi}^{\dag }(X^{\prime })\gamma ^{0}\left\vert 0,\mathrm{in}%
\right\rangle c_{v}^{-1},\;c_{v}=\langle 0,\mathrm{out}|0,\mathrm{in}\rangle
,  \notag \\
& S_{\text{\textrm{in}}}^{c}(X,X^{\prime })=i\left\langle 0,\mathrm{in}%
\right\vert \hat{T}\hat{\Psi}(X)\hat{\Psi}^{\dag }(X^{\prime })\gamma
^{0}\left\vert 0,\mathrm{in}\right\rangle ,  \notag \\
& S_{\text{\textrm{out}}}^{c}(X,X^{\prime })=i\left\langle 0,\mathrm{out}%
\right\vert \hat{T}\hat{\Psi}(X)\hat{\Psi}^{\dag }(X^{\prime })\gamma
^{0}\left\vert 0,\mathrm{out}\right\rangle ,  \label{m5.1}
\end{align}%
$\hat{T}$ stands for the chronological ordering operation, $P_{\mu
}=i\partial _{\mu }+eA_{\mu }(X)$, $\gamma ^{\mu }$ are the $\gamma $%
-matrices in $d$ dimensions, and the Dirac Heisenberg operator $\hat{\Psi}%
\left( X\right) $ is assigned to the Dirac field $\psi \left( X\right) $
according to conventional QED; see, e.g., \cite{GitTyu90}. It satisfies the
equal time canonical anticommutation relations%
\begin{equation}
\left. \left[ \hat{\Psi}\left( X\right) ,\hat{\Psi}\left( X^{\prime }\right) %
\right] _{+}\right\vert _{t=t^{\prime }}=0,\ \ \left. \left[ \hat{\Psi}%
\left( X\right) ,\hat{\Psi}^{\dagger }\left( X^{\prime }\right) \right]
_{+}\right\vert _{t=t^{\prime }}=\delta \left( \mathbf{r-r}^{\prime }\right)
.  \label{3.3}
\end{equation}%
In the general case the introduced matrix elements depend on the coordinate $%
x$ and thus are quasilocal quantities{\large .} Calculating these matrix
elements, we meet divergences that indicate the need to use a certain
regularization. In the present article we complement the previous
constructions presented in Ref. \cite{x-case}, formulating regularization
and renormalization procedures for calculating such quasilocal quantities.
We express the vector of the electric current density and the EMT of the
created particles via properly calculated corresponding vacuum means and
consider some important applications of these results.{\large \ }We show
that the smallness of the backreaction implies less restrictive conditions
on the electric field in QED with $x$-steps than those that exist for an
uniform field{\large . }The article is organized as follows:

In Sect. \ref{S3}, we recall briefly some basic facts of QED with $x$-steps
focusing on details important for treating the above mentioned problem\emph{.%
} We consider such local quantities as vacuum means of operators of the
electric current and of the energy--momentum tensor (EMT). These means
should be properly regularized and normalized. We propose a new
renormalization and volume regularization{\large \ }procedures which allow
one to calculate and distinguish physical parts of different matrix elements
of operators of the current and of the energy--momentum tensor, at the same
time relating the latter with characteristics of the vacuum instability.%
{\large \ } In Sect. \ref{S4}, using properly calculated vacuum mean values
of electric current and EMT, we discuss the backreaction problem. Creating
pairs from the vacuum the external electric field is losing its energy and
should be depleted over time. Thus, the applicability of the constant field
approximation is limited by the smallness of the backreaction. We study the
backreaction problem considering the case of an uniform electric field
confined between two capacitor plates separated by a finite distance $L$. We
compare the obtained restrictions with the consistency conditions derived by
us in Ref. \cite{GavG08}\ for QED with $t$-steps. To make text more readable
some features of the solutions of the Dirac equation with critical $x$-steps
are put in Appendix \ref{Ap}.

We use the relativistic units $\hslash =c=1$\ in which the fine structure
constant is $\alpha =e^{2}/c\hslash =e^{2}$.

\section{Means values\label{S3}}

\subsection{Regularization}

The basic mathematical tool of the standard strong-field QED consists of the
complete sets of exact solutions of relativistic wave equations
orthonormalized on a $t$ constant\ hyperplane.{\large \ }In the framework of%
{\large \ }the nonperturbative techniques elaborated in Ref. \cite{x-case}
for the strong-field QED with an $x$-step, which acts during the macroscopic
time{\large \ }$T$, one can construct such sets with the help of exact
solutions of the Dirac equation with an $x$-step.{\large \ }Formally, this
technique allows one to express{\large \ }propagators (\ref{m5.1}) via sums
over these solutions and obtain explicitly the vacuum matrix elements of
operators of the current and of the EMT,{\large \ }defined by Eq. (\ref%
{m5.3b}).{\large \ }These solutions are known in the form of the stationary
plane waves with given real longitudinal momenta{\large \ }$p^{\mathrm{L}}$%
{\large \ }and{\large \ }$p^{\mathrm{R}}${\large \ }in the regions $S_{%
\mathrm{L}}$\ and $S_{\mathrm{R}}$,%
\begin{eqnarray}
&&_{\;\zeta }\psi _{n}\left( X\right) \sim \exp \left[ -ip_{0}t+i\mathbf{p}%
_{\perp }\mathbf{r}_{\perp }+ip^{\mathrm{L}}\left( x-x_{\mathrm{L}}\right) %
\right] ,\ x\in S_{\mathrm{L}},  \notag \\
&&^{\;\zeta }\psi _{n}\left( X\right) \sim \exp \left[ -ip_{0}t+i\mathbf{p}%
_{\perp }\mathbf{r}_{\perp }+ip^{\mathrm{R}}\left( x-x_{\mathrm{R}}\right) %
\right] ,\ x\in S_{\mathrm{R}},  \notag \\
&&p^{\mathrm{L}}=\zeta \sqrt{\left[ \pi _{0}\left( \mathrm{L}\right) \right]
^{2}-\pi _{\bot }^{2}},\ \ p^{\mathrm{R}}=\zeta \sqrt{\left[ \pi _{0}\left(
\mathrm{R}\right) \right] ^{2}-\pi _{\bot }^{2}},\ \zeta =\pm \ ,  \notag \\
&&\pi _{0}\left( \mathrm{L/R}\right) =p_{0}-U_{\mathrm{L/R}},\ \ \pi _{\bot
}=\sqrt{\mathbf{p}_{\bot }^{2}+m^{2}}.  \label{m3a}
\end{eqnarray}%
{\large \ }They are parametrized by a set of quantum numbers{\large \ }$%
n=(p_{0},p_{\bot },\sigma )${\large \ }where $p_{0}$\ stands for total
energy, $p_{\bot }$\ is the transversal momentum (the index $\perp $\ stands
for components of the momentum that are perpendicular to the electric
field), and $\sigma $\ is the spin polarization.{\large \ }In the case of
the critical $x$-steps if $2\pi _{\bot }\leq \Delta U\ $\ there exist five
ranges\ $\Omega _{k}$, $k=1,...,5$, of quantum numbers{\large \ }$n${\large %
, }%
\begin{eqnarray}
&&p_{0}\geq U_{\mathrm{R}}+\pi _{\bot }\Longleftrightarrow \pi _{0}\left(
\mathrm{R}\right) \geq \pi _{\bot }\ \mathrm{\;if\;}n\in \Omega _{1},  \notag
\\
&&U_{\mathrm{R}}-\pi _{\bot }<p_{0}<U_{\mathrm{R}}+\pi _{\bot }\ \mathrm{%
\;if\;}n\in \Omega _{2},  \notag \\
&&U_{\mathrm{L}}+\pi _{\bot }\leq p_{0}\leq U_{\mathrm{R}}-\pi _{\bot }%
\mathrm{\;if\;}n\in \Omega _{3},  \notag \\
&&U_{\mathrm{L}}-\pi _{\bot }<p_{0}<U_{\mathrm{L}}+\pi _{\bot }\ \mathrm{%
\;if\;}n\in \Omega _{4},  \notag \\
&&p_{0}\leq U_{\mathrm{L}}-\pi _{\bot }\Longleftrightarrow -\pi _{0}\left(
\mathrm{L}\right) \geq \pi _{\bot }\ \mathrm{\;if\;}n\in \Omega _{5}\ ,
\label{ranges}
\end{eqnarray}%
where the solutions $\ _{\zeta }\psi _{n}\left( X\right) $ and $\ ^{\zeta
}\psi _{n}\left( X\right) $ have similar forms and properties for a given $%
\pi _{\bot }${\large . }In the ranges{\large \ }$\Omega _{2}${\large \ }and%
{\large \ }$\Omega _{4}${\large \ }we deal with standing waves completed by
linear superpositions of solutions{\large \ }$_{\;\pm }\psi _{n}\left(
X\right) ${\large \ }and{\large \ }$^{\;\pm }\psi _{n}\left( X\right) $%
{\large \ }with currents that are equal in magnitude.{\large \ }The explicit
structure of the solutions $_{\;\zeta }\psi _{n}\left( X\right) $\ and $%
^{\;\zeta }\psi _{n}\left( X\right) $\ on the whole axis $x$\ is given by
Eq. (\ref{ap.2}) in Appendix \ref{Ap}.{\large \ }It is shown in the
framework of QED by using one-particle mean currents and the energy fluxes
that, depending on the asymptotic behavior on the regions $S_{\mathrm{L}}$\
and $S_{\mathrm{R}}$, the plane waves $_{\;\zeta }\psi _{n}\left( X\right) $%
\ and $^{\;\zeta }\psi _{n}\left( X\right) $\ are identified unambiguously
as the components of the initial and final wave packets of electrons and
positrons; see sections V and VII and Appendices C1 and C2 in Ref. \cite%
{x-case}.{\large \ }However, unlike the global quantities considered in Ref.
\cite{x-case}, the quasilocal quantities (\ref{m5.3b}) calculated with the
help of such plane waves depend on the volume regularization parameters.
That is why in the following we turn to a clarification of the physical
meaning of these parameters, refining our regularization construction.

Stationary plane waves of the type (\ref{m3a}) are usually used in potential
scattering theory, where they represent one-particle states with
corresponding conserved longitudinal currents.{\large \ }Such an
one-particle consideration is valid in all ranges $\Omega _{k}$, except the
range $\Omega _{3}$, where the many-particle quantum field consideration is
essential. Note that the range $\Omega _{3}$\ often is referred to as the
Klein zone.

According to a standard volume regularization, we believe that the system
under consideration is situated in a large space-time box that has a spatial
volume $V_{\bot }$\ of the $(d-1)$-dimensional hypersurface orthogonal to
the electric field direction $x$\ and the time dimension $T$. Both $V_{\bot
} $\ and $T$\ are macroscopically large.{\large \ }It is supposed that all
the solutions of the Dirac equation $\psi \left( X\right) $\ are periodic
under transitions from one box to another.{\large \ }Under these
suppositions, the matrix elements of a charge transfer across the area $%
V_{\bot }$ of\ the $x=\mathrm{const}$\ hyperplane during the time $T$, given
by Eq. (\ref{c3}) in the Appendix \ref{Ap}, do not depend on $x$\ and can be
used as an inner product on the hyperplane{\large \ }$x=\mathrm{const}$%
{\large . }The plane waves of the type (\ref{m3a}) are orthonormalized with
respect to the inner product (\ref{c3}). Unlike wave functions of a
one-particle theory, the vacuum vectors $\left\vert 0,\mathrm{in}%
\right\rangle $\ and $\left\vert 0,\mathrm{out}\right\rangle $\ are global
states that are determined over the causally related part of a space volume
at a given time instant.{\large \ }To determine these vacua and construct
the corresponding in and out states in an adequate Fock space we have to use
a time-independent inner product of solutions $\psi \left( X\right) $\ and\ $%
\psi ^{\prime }\left( X\right) $ of the Dirac equation{\large \ }with\ the
field\ $E_{\mathrm{pristine}}\left( X\right) $\ on a $t$\ constant\
hyperplane.{\large \ }We recall that the periodic conditions are not imposed
in the $x$ direction{\large . }That is why, in contrast{\large \ }to the $t$%
-step case, the motion of particles in the $x$\ direction is unlimited.
However,{\large \ }it should be borne in mind that the electric field in
question is located inside the region $S_{\mathrm{int}}$\ during the time $T$%
. Consequently, causally related to the area $S_{\mathrm{int}}$ can there be
only such parts of the areas $S_{\mathrm{L}}$ and $S_{\mathrm{R}}$, which
are located from it at distances not exceeding $cT$.{\large \ }By virtue of
this, we propose to refine the volume regularization used in Ref. \cite%
{x-case}, defining the time-independent inner product on $t$\ constant\
hyperplane by%
\begin{equation}
\left( \psi ,\psi ^{\prime }\right) =\int_{V_{\bot }}d\mathbf{r}_{\bot
}\int\limits_{-K^{\left( \mathrm{L}\right) }}^{K^{\left( \mathrm{R}\right)
}}\psi ^{\dag }\left( X\right) \psi ^{\prime }\left( X\right) dx,  \label{t4}
\end{equation}%
where an integral over the spatial volume $V_{\bot }${\large \ }is completed
by an integral\ over the interval $\left[ K^{\left( \mathrm{L}\right)
},K^{\left( \mathrm{R}\right) }\right] $ in the $x$ direction. Here{\large \
}$K^{\left( \mathrm{L/R}\right) }${\large \ }are some macroscopic but finite%
{\large \ }parameters of the volume regularization, which are in spatial
areas where{\large \ }the electric field is absent,{\large \ }$\left\vert x_{%
\mathrm{FL}}\right\vert >K^{\left( \mathrm{L}\right) }\gg \left\vert x_{%
\mathrm{L}}\right\vert >0${\large \ }and $x_{\mathrm{FR}}>K^{\left( \mathrm{R%
}\right) }\gg x_{\mathrm{R}}>0${\large . }Such an inner product is
time-independent if solutions obey certain boundary conditions that allow
one to integrate by parts in Eq. (\ref{t4}) neglecting boundary terms. The
standard understanding as regards the volume regularization in QFT that
physical states are wave packets that vanish on the remote boundaries is
also accepted. The inner product (\ref{t4}) is conserved for such states.
However, considering solutions of the type (\ref{m3a}) that are plane wave
components of the packets, which do not vanish at the spatial infinity, we
must accept some additional assumptions to provide the time independence of
the inner product (\ref{t4}). First we note that{\large \ }states with
different quantum numbers $n$\ are independent, therefore decompositions of
vacuum matrix elements (\ref{m5.3b}) into the plane waves with given $n$\ do
not contain interference terms; see Appendix \ref{Ap} for details.{\large \ }%
That is why it is enough to consider Eq. (\ref{t4}) only for a particular
case of plane waves{\large \ }$_{\;\zeta }\psi _{n}\left( X\right) ${\large %
\ }and{\large \ }$^{\;\zeta }\psi _{n}\left( X\right) ${\large \ }with equal%
{\large \ }$n$. Assuming that the areas $S_{\mathrm{L}}$\ and $S_{\mathrm{R}%
} $\ are much wider than the area{\large \ }$S_{\mathrm{int}}${\large , }%
\begin{equation}
K^{\left( \mathrm{L}\right) }-\left\vert x_{\mathrm{L}}\right\vert
,K^{\left( \mathrm{R}\right) }-x_{\mathrm{R}}\gg x_{\mathrm{R}}-x_{\mathrm{L}%
},  \label{m0}
\end{equation}%
and the potential energy $U(x)$\ is a smooth function,{\large \ }the
principal value of the integral (\ref{t4}) is determined by integrals over
the areas $x\in \left[ -K^{\left( \mathrm{L}\right) },x_{\mathrm{L}}\right] $
and $x\in \left[ x_{\mathrm{R}},K^{\left( \mathrm{R}\right) }\right] $,
where the electric field is zero. Thus, it is possible to evaluate integrals
of the form (\ref{t4}) for any form of the external field, using only the
asymptotic behavior (\ref{m3a}) of functions in the regions $S_{\mathrm{L}}$%
and $S_{\mathrm{R}}$ \ where an electric field is absent and particles are
free. The form of the electric field affects only the coefficients $g$
entering into the mutual decompositions of the solutions,
\begin{eqnarray}
\eta _{\mathrm{L}}\ ^{\zeta }\psi _{n}\left( X\right) &=&\ _{+}\psi
_{n}\left( X\right) g\left( _{+}\left\vert ^{\zeta }\right. \right) -\
_{-}\psi _{n}\left( X\right) g\left( _{-}\left\vert ^{\zeta }\right. \right)
,  \notag \\
\eta _{\mathrm{R}}\ _{\zeta }\psi _{n}\left( X\right) &=&\ ^{+}\psi
_{n}\left( X\right) g\left( ^{+}\left\vert _{\zeta }\right. \right) -\
^{-}\psi _{n}\left( X\right) g\left( ^{-}\left\vert _{\zeta }\right. \right)
,  \label{rel1}
\end{eqnarray}%
here $\eta _{\mathrm{L}}=\eta _{\mathrm{R}}=1$ for $n\in \Omega _{1}$, $\eta
_{\mathrm{L}}=\eta _{\mathrm{R}}=-1$ for $n\in \Omega _{5},$ and $\eta _{%
\mathrm{L}}=-\eta _{\mathrm{R}}=1$ for $n\in \Omega _{3}$. The coefficients $%
g^{\prime }$s are defined by Eq. (\ref{c12}) and satisfy the unitary
relations (\ref{UR}) represented in Appendix \ref{Ap}. One can see that the
norms of the plane waves $_{\;\zeta }\psi _{n}\left( X\right) ${\large \ }and%
{\large \ }$^{\;\zeta }\psi _{n}\left( X\right) $ with respect to the inner
product (\ref{t4}) are proportional to the macroscopically large parameters%
{\large \ }$\tau ^{\left( \mathrm{L}\right) }=K^{\left( \mathrm{L}\right)
}/v^{\mathrm{L}}${\large \ }and{\large \ }$\tau ^{\left( \mathrm{R}\right)
}=K^{\left( \mathrm{R}\right) }/v^{\mathrm{R}}${\large , }where $v^{\mathrm{L%
}}=\left\vert p^{\mathrm{L}}/\pi _{0}\left( \mathrm{L}\right) \right\vert >0$
and $v^{\mathrm{R}}=\left\vert p^{\mathrm{R}}/\pi _{0}\left( \mathrm{R}%
\right) \right\vert >0$ are absolute values of longitudinal velocities of
particles in the regions $S_{\mathrm{L}}${\large \ }and{\large \ }$S_{%
\mathrm{R}}$, respectively; see Sect. IIIC.2 and Appendix B in Ref. \cite%
{x-case} for details.

It is shown (see Appendix B in Ref. \cite{x-case}) that\ the following
couples of plane waves are orthogonal with \ respect to the inner product (%
\ref{t4})%
\begin{equation}
\left( _{\zeta }\psi _{n},^{-\zeta }\psi _{n}\right) =0,\ \ n\in \Omega
_{1}\cup \Omega _{5}\ ;\ \ \left( _{\zeta }\psi _{n},^{\zeta }\psi
_{n}\right) =0,\ \ n\in \Omega _{3}\ ,  \label{i7}
\end{equation}%
if the parameters of the volume regularization $\tau ^{\left( \mathrm{L/R}%
\right) }$\ satisfy the condition%
\begin{equation}
\tau ^{\left( \mathrm{L}\right) }-\tau ^{\left( \mathrm{R}\right) }=O\left(
1\right) ,  \label{i8}
\end{equation}%
where $O\left( 1\right) $ denotes terms that are negligibly small in
comparison with the macroscopic quantities $\tau ^{\left( \mathrm{L/R}%
\right) }${\large . }One can see that $\tau ^{\left( \mathrm{L}\right) }$
and $\tau ^{\left( \mathrm{R}\right) }$ are macroscopic times of motion of
electrons and positrons in the areas $S_{\mathrm{L}}$\ and $S_{\mathrm{R}}$
respectively and they are equal,{\large \ }%
\begin{equation}
\tau ^{\left( \mathrm{L}\right) }=\tau ^{\left( \mathrm{R}\right) }=\tau .
\label{m4}
\end{equation}%
It allows one to introduce an unique time of motion $\tau $\ for all the
particles in the system under consideration.{\large \ }This time can be
interpreted as a time of observation of the evolution of the system under
consideration.

Thus, there are constructed two linearly independent couples of complete%
{\large \ }on the $t$-constant hyperplane states with given $n${\large \ }%
that are either "\textrm{in}" or "\textrm{out}" states according to the
physical interpretation given in Ref. \cite{x-case}),%
\begin{eqnarray}
&&\mathrm{in:}\mathrm{\ }\psi _{n_{1}}^{\left( \mathrm{in},+\right) }=\
_{+}\psi _{n_{1}},\ \psi _{n_{1}}^{\left( \mathrm{in},-\right) }=\ ^{-}\psi
_{n_{1}};\;\psi _{n_{5}}^{\left( \mathrm{in},+\right) }=\ ^{+}\psi _{n_{5}},%
\mathrm{\ }\psi _{n_{5}}^{\left( \mathrm{in},-\right) }=\ _{-}\psi _{n_{5}};
\notag \\
&&\psi _{n_{3}}^{\left( \mathrm{in},+\right) }=\ ^{-}\psi _{n_{3}},\ \ \psi
_{n_{3}}^{\left( \mathrm{in},-\right) }=\ _{-}\psi _{n_{3}};  \notag \\
&&\mathrm{out:}\mathrm{\ }\psi _{n_{1}}^{\left( \mathrm{out},-\right) }=\
_{-}\psi _{n_{1}},\ \psi _{n_{1}}^{\left( \mathrm{out},+\right) }=\ ^{+}\psi
_{n_{1}};\;\psi _{n_{5}}^{\left( \mathrm{out},-\right) }=\ ^{-}\psi _{n_{5}},%
\mathrm{\ }\psi _{n_{5}}^{\left( \mathrm{out},+\right) }=\ _{+}\psi
_{n_{5}};\ \   \notag \\
&&\psi _{n_{3}}^{\left( \mathrm{out},+\right) }=\ ^{+}\psi _{n_{3}}\ \ \
\psi _{n_{3}}^{\left( \mathrm{out},-\right) }=\ _{+}\psi _{n_{3}},\;n_{k}\in
\Omega _{k}.  \label{in-out}
\end{eqnarray}%
{\large \ }

Under condition (\ref{i8}) the norms of the plane waves on the $t$-constant
hyperplane{\large \ are}%
\begin{eqnarray}
&&\left( \ _{\zeta }\psi _{n},\ _{\zeta }\psi _{n}\right) =\left( \ ^{\zeta
}\psi _{n},\ ^{\zeta }\psi _{n}\right) =\mathcal{M}_{n}\ \mathrm{if}\ \ n\in
\Omega _{1}\cup \Omega _{3}\cup \Omega _{5},  \notag \\
&&\mathcal{M}_{n}=2\frac{\tau }{T}\left\vert g\left( _{+}\left\vert
^{+}\right. \right) \right\vert ^{2}\ \ \mathrm{if}\ \ n\in \Omega _{1}\cup
\Omega _{5},  \notag \\
&&\mathcal{M}_{n}=2\frac{\tau }{T}\left\vert g\left( _{+}\left\vert
^{-}\right. \right) \right\vert ^{2}\ \ \mathrm{if}\ \ n\in \Omega _{3},
\label{i12}
\end{eqnarray}%
where the coefficients $g$ are defined by Eq. (\ref{c12}) in Appendix \ \ref%
{Ap}.

Note that the plane waves {\large \ }$_{\;\zeta }\psi _{n}\left( X\right) $%
{\large \ }and{\large \ }$^{\;\zeta }\psi _{n}\left( X\right) $ coincide
with solutions of the Dirac equation, unknown in an explicit form, with the
field{\large \ }$E_{\mathrm{pristine}}\left( X\right) ${\large \ }for{\large %
\ }$t_{\mathrm{in}}<t<t_{\mathrm{out}}${\large . }The time independence of
the inner product (\ref{t4}){\large \ }implies that orthonormality relations
for the plane waves{\large \ }$_{\;\zeta }\psi _{n}\left( X\right) ${\large %
\ }and{\large \ }$^{\;\zeta }\psi _{n}\left( X\right) $ on{\large \ }the $t$%
-constant{\large \ }hyperplane{\large \ }coincide in fact with
orthonormality relations for the solutions of the Dirac equation with the
field\ $E_{\mathrm{pristine}}\left( X\right) $\ under\ the corresponding
initial or final conditions.{\large \ }In particular,{\large \ }one can
construct sets of linearly independent plane waves related to either initial
or final particles as follows.

We believe that the sets (\ref{in-out}) together with solutions $\psi
_{n_{2}}$ and $\psi _{n_{4}}$, which are not essential for the problems
discussed in this article, form complete \textrm{in} and \textrm{out} sets
in the Hilbert space of Dirac spinors at any fixed time instant $t$.{\large %
\ }This assumption is equivalent to the existence of the propagation
function in the space of solutions, which satisfies the standard equal time
boundary condition. This fact is crucial for a nonperturbative formulation
of QED with $x$-steps.

For a convenience of the reader, we recall some basic points of this
formulation.{\large \ }One can decompose the Heisenberg operator $\hat{\Psi}%
\left( X\right) $\ into the solutions of either initial or final complete
sets (\ref{in-out}) and construct in this way \textrm{in-} and \textrm{out}%
-states in an adequate Fock space:%
\begin{eqnarray}
&&\hat{\Psi}\left( X\right) =\sum_{n}\mathcal{M}_{n}^{-1/2}\left[ \ A_{n}(%
\mathrm{in})\ \psi _{n}^{\left( \mathrm{in},+\right) }\left( X\right) +\
B_{n}^{\dagger }(\mathrm{in})\ \psi _{n}^{\left( \mathrm{in},-\right)
}\left( X\right) \right]  \notag \\
&&\ =\sum_{n}\mathcal{M}_{n}^{-1/2}\left[ \ A_{n}(\mathrm{out})\ \psi
_{n}^{\left( \mathrm{out},+\right) }\left( X\right) +\ B_{n}^{\dagger }(%
\mathrm{out})\ \psi _{n}^{\left( \mathrm{out},-\right) }\left( X\right) %
\right] ,  \label{dec}
\end{eqnarray}%
where operator-valued coefficients are determined with the help of the inner
product (\ref{t4}),
\begin{eqnarray*}
&&A_{n}(\mathrm{in})=\left( \psi _{n}^{\left( \mathrm{in},+\right) },\hat{%
\Psi}\right) ,\;B_{n}^{\dagger }(\mathrm{in})=\left( \psi _{n}^{\left(
\mathrm{in},-\right) },\hat{\Psi}\right) ; \\
&&A_{n}(\mathrm{out})=\left( \psi _{n}^{\left( \mathrm{out},+\right) },\hat{%
\Psi}\right) ,\;B_{n}^{\dagger }(\mathrm{out})=\left( \psi _{n}^{\left(
\mathrm{out},-\right) },\hat{\Psi}\right) .
\end{eqnarray*}%
These operators{\Huge \ }define{\Huge \ }Fermi annihilation and creation
operators for initial or final particles as follows:
\begin{eqnarray}
A_{n_{1}}(\mathrm{in}) &=&\ _{+}a_{n_{1}}(\mathrm{in}),\;B_{n_{1}}^{\dagger
}(\mathrm{in})=\ ^{-}a_{n_{1}}(\mathrm{in});\;A_{n_{1}}(\mathrm{out})=\
^{+}a_{n_{1}}(\mathrm{out})\ ,\;B_{n_{1}}^{\dagger }(\mathrm{out})=\
_{-}a_{n_{1}}(\mathrm{out})\ ;  \notag \\
A_{n_{2}}(\mathrm{in}) &=&A_{n_{2}}(\mathrm{out})=a_{n_{2}},\;B_{n_{2}}^{%
\dagger }(\mathrm{in})=B_{n_{2}}^{\dagger }(\mathrm{out})=0\;;  \notag \\
A_{n_{3}}(\mathrm{in}) &=&\ ^{-}a_{n_{3}}(\mathrm{in}),\;B_{n_{3}}^{\dagger
}(\mathrm{in})=\ _{-}b_{n_{3}}^{\dagger }(\mathrm{in});\;A_{n_{3}}(\mathrm{%
out})=\ ^{+}a_{n_{3}}(\mathrm{out}),\;B_{n_{3}}^{\dagger }(\mathrm{out})=\
_{+}b_{n_{3}}^{\dagger }(\mathrm{out});  \notag \\
A_{n_{4}}(\mathrm{in}) &=&A_{n_{4}}(\mathrm{out})=0,\;B_{n_{4}}^{\dagger }(%
\mathrm{in})=B_{n_{4}}^{\dagger }(\mathrm{out})=b_{n}^{\dagger }\;;  \notag
\\
A_{n_{5}}(\mathrm{in}) &=&\;^{+}b_{n_{5}}^{\dag }(\mathrm{in}%
),\;B_{n_{5}}^{\dagger }(\mathrm{in})=\ _{-}b_{n_{5}}^{\dag }(\mathrm{in}%
);\;A_{n_{5}}(\mathrm{out})=_{+}b_{n_{5}}^{\dag }(\mathrm{out}%
),\;B_{n_{5}}^{\dagger }(\mathrm{out})=\ ^{-}b_{n_{5}}^{\dag }(\mathrm{out}).
\label{ab}
\end{eqnarray}%
All $a$ and $b\ $are annihilation and all $a^{\dag }$ and $b^{\dag }$ are
creation operators, whereas all $a$ and $a^{\dag }$ describe electrons and
all $b$ and $b^{\dag }$ describe positrons. All the operators labeled by the
argument "\textrm{in"} are interpreted\ as \textrm{in}-operators, whereas
all the operators labeled by the argument "\textrm{out" }are interpreted as
\textrm{out}-operators. This identification is confirmed by a detailed
analysis in Sects. V,VI, and VII of Ref. \cite{x-case}. Equations (\ref{3.3}%
) yield the standard anticommutation rules for the creation and annihilation
\textrm{in}- or \textrm{out-}operators introduced . The unitary
transformation (\ref{rel1}) implies canonical transformations between the
\textrm{in} and \textrm{out}-operators.

The two vacuum vectors $\left\vert 0,\mathrm{in}\right\rangle $ and $%
\left\vert 0,\mathrm{out}\right\rangle $ are null vectors for all $a$ and $%
b\ $annihilation operators given by Eq. (\ref{ab}). The partial vacua in the
Fock subspaces with given{\large \ }$n${\large \ }are stable only in $\Omega
_{i},$ $i=1,2,4,5$. The electric field violates the vacuum stability in the
range $\Omega _{3}$; thus, pair creation occurs only in this range and%
{\large \ }the total vacuum-to-vacuum transition amplitude $c_{v}$ is formed
due to the latter instability. Differential mean numbers of electrons $%
N_{n}^{a}\left( \mathrm{out}\right) $ and positrons $N_{n}^{b}\left( \mathrm{%
out}\right) $,\ $n\in \Omega _{3},$ created from the vacuum are equal, $%
N_{n}^{b}\left( \mathrm{out}\right) =N_{n}^{a}\left( \mathrm{out}\right)
=N_{n}^{\mathrm{cr}}$, and have the forms%
\begin{eqnarray}
&&N_{n}^{a}\left( \mathrm{out}\right) =\left\langle 0,\mathrm{in}\left\vert
\ ^{+}a_{n}^{\dagger }(\mathrm{out})\ ^{+}a_{n}(\mathrm{out})\right\vert 0,%
\mathrm{in}\right\rangle =\left\vert g\left( _{-}\left\vert ^{+}\right.
\right) \right\vert ^{-2},  \notag \\
&&N_{n}^{b}\left( \mathrm{out}\right) =\left\langle 0,\mathrm{in}\left\vert
\ _{+}b_{n}^{\dagger }(\mathrm{out})\ _{+}b_{n}(\mathrm{out})\right\vert 0,%
\mathrm{in}\right\rangle =\left\vert g\left( _{+}\left\vert ^{-}\right.
\right) \right\vert ^{-2}.  \label{Na}
\end{eqnarray}%
The total number of pairs\ created from the vacuum reads%
\begin{equation}
N^{\mathrm{cr}}=\sum_{n\in \Omega _{3}}N_{n}^{\mathrm{cr}}=\sum_{n\in \Omega
_{3}}\left\vert g\left( _{-}\left\vert ^{+}\right. \right) \right\vert ^{-2}.
\label{TN}
\end{equation}

A detailed consideration of various physical processes (see Sects. V, VI,
and VII and Appendices C and D in Ref. \cite{x-case}) shows that in the
range $\Omega _{1}$\ there exist only \textrm{in}- and \textrm{out}%
-electrons, whereas in the range $\Omega _{5}$\ there exist only \textrm{in}%
- and \textrm{out}-positrons. In these ranges electrons and positrons are
subject to scattering and reflection only. In the range $\Omega _{2}$\ there
exist only electrons that are subjected to the total\ reflection and have an
unbounded motion in the $x\rightarrow -\infty $\ direction. In the range $%
\Omega _{4}$\ there exist only positrons that are also subject to total\
reflection but have an unbounded motion in the $x\rightarrow +\infty $\
direction. No particle creation in these four ranges is possible. We note
that all the rules of the potential scattering theory can be derived in the
framework of QFT.

In the range $\Omega _{3}$\ the quantum field description of the processes
is essential. In this range there exist \textrm{in}- and \textrm{out}%
-electrons (localizable electron wave packets) that can be situated only to
the left of the step, and in- and out-positrons (localizable positron wave
packets) that can be situated only to the right of the step. In this range,
all the partial vacua with given $n$ are unstable, and particle production
from vacuum is possible. These pairs consist of out-electrons and
out-positrons that appear on the left and on the right of the step and move
there\ to the left and to the right, respectively. At the same time, the
in-electrons that move\ to the step from the left are subject to the total
reflection. After\ being reflected they move to the left of the step already
as out-electrons. Similarly, the \textrm{in}-positrons that move\ to the
step from the right are subject to the total reflection.\ After being
reflected they move to the right of the step already as \textrm{out}%
-positrons.

In the following the effect of pair production from vacuum is most
interesting. In this case{\large \ }electrons and positrons leaving the area
$S_{\mathrm{int}}$ enter the areas $S_{\mathrm{L}}$ and $S_{\mathrm{R}}$
respectively. There they continue to move in opposite directions with
constant longitudinal velocities $-v^{\mathrm{L}}$ and $v^{\mathrm{R}}$. Then%
{\large \ }in the range{\large \ }$\Omega _{3}$ the parameter $\tau $\ can
be interpreted as the observation time of the pair production process.

\subsection{Renormalization}

Here we complement the above consideration, indicating a procedure that
allows one to link quasilocal quantities (\ref{m5.3b}) with observable
physical quantities specifying the vacuum instability.{\large \ }In the
general case, the matrix elements (\ref{m5.3b}){\large \ }contain local
contributions due to the vacuum polarization and contributions due to the
vacuum instability caused by the external field for all the time $T=t_{%
\mathrm{out}}-t_{\mathrm{in}}$ of his action.{\large \ }We believe that
under the condition (\ref{m0}) all local contributions due to the existence
of the external field in the area\ $S_{\mathrm{int}}$\ can be neglected.%
{\large \ }Therefore, it is enough to know lthe ongitudinal currents and
energy fluxes through the surfaces\ $x=x_{\mathrm{L}}$\ and\ $x=x_{\mathrm{R}%
}$ to construct the initial and final states and to find relations between
them and the characteristics of the vacuum instability.{\large \ }It is
clear that such fluxes of created pairs depend on the parameter of the
volume regularization $\tau $\ due to the presence of the normalization
factor $M_{n}^{-1/2}$\ in the field operator decomposition (\ref{dec}).%
{\large \ }Thus, we can find their relation to observable physical
quantities and obtain a relation between the parameter $\tau $\ and the
whole time $T$.{\large \ }Such a relation fixes the\ proposed
renormalization procedure.

Using the decompositions (\ref{dec}) and definitions of the vacua $%
\left\vert 0,\mathrm{in}\right\rangle $\ and $\left\vert 0,\mathrm{out}%
\right\rangle ,$\ one can find an explicit form of singular functions (\ref%
{m5.1}) for all the ranges $\Omega _{k}$;{\large \ }see Sect. VIII in Ref.
\cite{x-case} for details.{\large \ }However, since the partial vacua are
stable in the ranges $\Omega _{k}$, $k=1,2,4,5,$ the corresponding
contributions from the propagators $S^{c}$, $S_{\text{\textrm{in}}}^{c}$,
and $S_{\text{\textrm{out}}}^{c}$\ to the vacuum matrix elements{\large \ }%
coincide. That is why we do not need explicit forms of the singular
functions\ in these ranges.{\large \ }It is enough to note that{\large \ }%
using asymptotic behavior of the solutions $_{\;\zeta }\psi _{n}\left(
X\right) $\ and $^{\;\zeta }\psi _{n}\left( X\right) $ one can verify that
contributions due to these ranges to vacuum currents and energy fluxes in
the areas $S_{\mathrm{L}}$\ and $S_{\mathrm{R}}$\ are absent. Similar
contributions to the diagonal elements $\left\langle T_{\mu \mu
}(x)\right\rangle ^{c}$\ do not depend on the electric field\textbf{\ }and
have to be neglected according to the standard renormalization procedure.%
{\large \ }Therefore, contributions due to the propagator $S^{c}$\ can
affect only the vacuum polarization in the field region $S_{\mathrm{int}}$,
the formal substantiation of this fact is not directly related to the issues
under consideration and will be considered in a future publication.

We are interested in vacuum instability effects, which are formed
exclusively in the Klein zone $\Omega _{3}$. As was mentioned earlier, under
condition (\ref{m0}) the principal value of integral (\ref{t4}) is
determined by integrals over the areas $x\in \left[ -K^{\left( \mathrm{L}%
\right) },x_{\mathrm{L}}\right] $ and $x\in \left[ x_{\mathrm{R}},K^{\left(
\mathrm{R}\right) }\right] $, where the electric field is zero. This implies
that only the areas $S_{\mathrm{L}}$ and $S_{\mathrm{R}}$ determine
contributions of the vacuum instability effects to matrix elements (\ref%
{m5.3b}). For this reason, in what follows, we consider these matrix elements%
{\large \ }only in the areas $S_{\mathrm{L}}$ and $S_{\mathrm{R}}$,

Thus, we may limit ourselves to the contributions to the singular functions
that are formed in the Klein zone (we denote such contributions with a tilde
at the top),%
\begin{eqnarray}
&&\tilde{S}^{c}(X,X^{\prime })=\theta (t-t^{\prime })\,\tilde{S}^{-}\left(
X,X^{\prime }\right) -\theta (t^{\prime }-t)\,\tilde{S}^{+}\left(
X,X^{\prime }\right) ,  \notag \\
&&\tilde{S}^{-}(X,X^{\prime })=i\sum_{n\in \Omega _{3}}\mathcal{M}_{n}^{-1}\
^{+}\psi _{n}\left( X\right) w_{n}\left( +|+\right) \ ^{-}\bar{\psi}%
_{n}\left( X^{\prime }\right) ,  \notag \\
&&\tilde{S}^{+}(X,X^{\prime })=i\sum_{n\in \Omega _{3}}\mathcal{M}_{n}^{-1}\
_{-}\psi _{n}\left( X\right) w_{n}\left( -|-\right) \ _{+}\bar{\psi}%
_{n}\left( X^{\prime }\right) ;  \label{m5.2} \\
&&\tilde{S}_{\mathrm{in/out}}^{c}(X,X^{\prime })=\theta (t-t^{\prime })%
\tilde{S}_{\mathrm{in/out}}^{-}(X,X^{\prime })-\theta (t^{\prime }-t)\tilde{S%
}_{\mathrm{in/out}}^{+}(X,X^{\prime })\,,  \notag \\
&&\tilde{S}_{\mathrm{in/out}}^{-}(X,X^{\prime })=i\sum_{n\in \Omega _{3}}%
\mathcal{M}_{n}^{-1}\ ^{\mp }\psi _{n}\left( X\right) \ ^{\mp }\bar{\psi}%
_{n}\left( X^{\prime }\right) ,  \notag \\
&&\tilde{S}_{\mathrm{in/out}}^{+}(X,X^{\prime })=i\sum_{n\in \Omega _{3}}%
\mathcal{M}_{n}^{-1}\ _{\mp }\psi _{n}\left( X\right) \ _{\mp }\bar{\psi}%
_{n}\left( X^{\prime }\right) ,  \label{m5.4}
\end{eqnarray}%
where $\bar{\psi}_{n}\left( X\right) =\psi _{n}^{\dag }\left( X\right)
\gamma ^{0}$ , and the amplitudes $w_{n}$ are
\begin{eqnarray}
&&w_{n}\left( +|+\right) =g\left( ^{+}\left\vert _{-}\right. \right) g\left(
^{-}\left\vert _{-}\right. \right) ^{-1}=g\left( _{+}\left\vert ^{-}\right.
\right) g\left( _{+}\left\vert ^{+}\right. \right) ^{-1},  \notag \\
&&w_{n}\left( -|-\right) =g\left( ^{-}\left\vert _{+}\right. \right) g\left(
^{-}\left\vert _{-}\right. \right) ^{-1}=g\left( _{-}\left\vert ^{+}\right.
\right) g\left( _{+}\left\vert ^{+}\right. \right) ^{-1}.  \label{m14}
\end{eqnarray}%
Here $g^{\prime }$s are given by Eq. (\ref{c12}) in Appendix \ref{Ap}.
Taking into account Eqs. (\ref{i12}) and Eq. (\ref{Na}) one can see that the
normalization factor $M_{n}^{-1}$\ depends on the differential numbers $%
N_{n}^{\mathrm{cr}}$,\ and on the ratio{\large \ }$T/\tau ${\large ,}%
\begin{equation}
\mathcal{M}_{n}^{-1}=\frac{T}{2\tau }N_{n}^{\mathrm{cr}}\ \mathrm{for}\ \
n\in \Omega _{3}.  \label{m14b}
\end{equation}%
Then means (\ref{m5.3b}) can be expressed via singular functions (\ref{m5.4}%
) as follows:%
\begin{align}
& \left\langle J^{\mu }(x)\right\rangle _{\mathrm{in/out}}=-\frac{ie}{2}%
\text{\textrm{tr}}\left\{ \gamma ^{\mu }\left[ \tilde{S}_{\mathrm{in/out}%
}^{-}(X,X^{\prime })-\tilde{S}_{\mathrm{in/out}}^{+}(X,X^{\prime })\right]
\right\} |_{X=X^{\prime }}\ ,  \notag \\
& \left\langle T_{\mu \nu }(x)\right\rangle _{\mathrm{in/out}}=\frac{i}{2}%
\text{\textrm{tr}}\left\{ A_{\mu \nu }\left[ \tilde{S}_{\mathrm{in/out}%
}^{-}(X,X^{\prime })-\tilde{S}_{\mathrm{in/out}}^{+}(X,X^{\prime })\right]
\right\} |_{X=X^{\prime }}\ .  \label{m5.5}
\end{align}

It follows from Eq. (\ref{m5.5}) that due to the cylindrical symmetry of a
problem the transversal components of the mean currents and non-diagonal
components of mean values of the EMT vanish,%
\begin{eqnarray}
&&\left\langle J^{k}(x)\right\rangle _{\mathrm{in}}=\left\langle
J^{k}(x)\right\rangle _{\mathrm{out}}=\left\langle J^{k}(x)\right\rangle
^{c}=0,\ \ \left\langle T_{0k}(x)\right\rangle _{\mathrm{in}}=\left\langle
T_{0k}(x)\right\rangle _{\mathrm{out}}\ =\left\langle T_{0k}(x)\right\rangle
^{c}=0,\ \ k\neq 1,  \notag \\
&&\left\langle T_{lk}(x)\right\rangle _{\mathrm{in}}=\left\langle
T_{lk}(x)\right\rangle _{\mathrm{out}}=\left\langle T_{lk}(x)\right\rangle
^{c}=0,\ \ k\neq l,\ k,l=2,\ldots .  \label{m6}
\end{eqnarray}

The relations between \textrm{in--out} matrix elements and \textrm{in--in}
and \textrm{out--out} means can be written in terms of a singular function $%
S^{p}(X,X^{\prime }),$
\begin{equation}
S^{p}(X,X^{\prime })=S_{\text{\textrm{in}}}(X,X^{\prime })-S^{c}(X,X^{\prime
}),  \label{Sp}
\end{equation}%
which was found in Ref. \cite{x-case},%
\begin{equation}
S^{p}(X,X^{\prime })=i\sum_{n\in \Omega _{3}}\mathcal{M}%
_{n}^{-1}g(^{-}|_{-})^{-1}\ _{-}\psi _{n}\left( X\right) \ ^{-}\bar{\psi}%
_{n}\left( X^{\prime }\right) .  \label{m13.2}
\end{equation}%
These relations have the form%
\begin{equation}
\left\langle J^{\mu }(x)\right\rangle _{\mathrm{in}}=\left\langle J^{\mu
}(x)\right\rangle ^{c}+\left\langle J^{\mu }(x)\right\rangle ^{p},\ \
\left\langle T_{\mu \nu }(x)\right\rangle _{\mathrm{in}}=\ \left\langle
T_{\mu \upsilon }(x)\right\rangle ^{c}+\ \left\langle T_{\mu \upsilon
}(x)\right\rangle ^{p},  \label{m13.3}
\end{equation}%
where
\begin{equation}
\left\langle J^{\mu }(x)\right\rangle ^{p}=-ie\text{\textrm{tr}}\left[
\gamma ^{\mu }S^{p}(X,X^{\prime })\right] |_{X=X^{\prime }},\ \ \left\langle
T_{\mu \upsilon }(x)\right\rangle ^{p}=i\text{\textrm{tr}}\left[ A_{\mu
\upsilon }S^{p}(X,X^{\prime })\right] |_{X=X^{\prime }}.  \label{m13.1}
\end{equation}%
As it follows from Eq. (\ref{m13.2}) the quantities (\ref{m13.1}) are formed
exclusively in the Klein zone $\Omega _{3}$. Besides, one can see from Eq. (%
\ref{m6}) that only the quantities $\left\langle J^{0}(x)\right\rangle ^{p}$%
, $\left\langle J^{1}(x)\right\rangle ^{p}$, and $\left\langle
T^{10}(x)\right\rangle ^{p}$ and$\ \left\langle T_{\mu \mu }(x)\right\rangle
^{p}$ are nonzero.

It would be interesting and practically important to associate the
quantities (\ref{m5.5}) with quantities characterizing directly the effect
of pair production. Below, this problem is solved for the first time.

It should be noted that in QED with $t$-steps \cite{Gitman} Heisenberg%
{\large \ }operators of physical quantities (for example, the kinetic energy
operator of the Dirac field) are obviously time dependent.{\large \ }That is
why one can determine contributions of the final particles, using \textrm{%
in-in} vacuum means, and setting{\large \ }$t\rightarrow \infty ${\large \ }%
(which means considering the time instant when the external field is already
switched off and all the corresponding effects of the vacuum polarization
vanish). In the case under consideration mean values are time independent,
therefore this possibility is not available. An alternative procedure is
required to determine contributions of the final particles.{\large \ }Such a
procedure is proposed below.

In the general case, the means (\ref{m5.5}) contain local contributions of
the vacuum polarization and contributions from the vacuum instability caused
by the external field during the whole time $T=t_{\mathrm{out}}-t_{\mathrm{in%
}}$ of its existence.{\large \ }As was already mentioned,{\large \ }due to
condition (\ref{m0}) all the local contributions due to the presence of an
external field in the area\ $S_{\mathrm{int}}$\ can be neglected.{\large \ }%
Therefore, we believe that it is enough to know the longitudinal currents
and the energy fluxes through the surfaces\ $x=x_{\mathrm{L}}$\ and\ $x=x_{%
\mathrm{R}}$ to evaluate the contributions of the initial and final states.

Let us consider nonzero longitudinal fluxes in the areas $S_{\mathrm{L}}$
and $S_{\mathrm{R}}$. Using Eq. (\ref{m3}) in singular functions (\ref{m5.4}%
), one can find%
\begin{eqnarray}
&&\left\langle J^{1}(x)\right\rangle _{\mathrm{in}}=-\ \left\langle
J^{1}(x)\right\rangle _{\mathrm{out}}=\left\langle J^{1}(x)\right\rangle
^{p}=\bar{J}^{1}\ ,\ \ x\in S_{\mathrm{L}}\;\mathrm{or}\ S_{\mathrm{R}},
\notag \\
&&\ \bar{J}^{1}=\frac{e}{2}\sum_{n\in \Omega _{3}}\tilde{j}_{n}^{1},\ \
\tilde{j}_{n}^{1}=N_{n}^{\mathrm{cr}}(\tau V_{\perp })^{-1};  \label{m7} \\
&&\left\langle T^{10}(x)\right\rangle _{\mathrm{in}}=-\ \left\langle
T^{10}(x)\right\rangle _{\mathrm{out}}=\left\langle T^{10}(x)\right\rangle
^{p},  \notag \\
&&\left\langle T^{10}(x)\right\rangle ^{p}=\left\{
\begin{tabular}{l}
$\bar{T}^{10}\left( \mathrm{L}\right) =-\frac{1}{2}\sum_{n\in \Omega
_{3}}\pi _{0}\left( \mathrm{L}\right) \tilde{j}_{n}^{1}\ ,\ \ x\in S_{%
\mathrm{L}}$ \\
$\bar{T}^{10}\left( \mathrm{R}\right) =\frac{1}{2}\sum_{n\in \Omega
_{3}}\left\vert \pi _{0}\left( \mathrm{R}\right) \right\vert \tilde{j}%
_{n}^{1}\ ,\ \ \ x\in S_{\mathrm{R}}$%
\end{tabular}%
\right. .  \label{m8}
\end{eqnarray}%
Note that the electric current densities (\ref{m7}) (as well as all
components $\tilde{j}_{n}^{1}$) are conserved along the axis $x$, while the
energy flux densities $\bar{T}^{10}\left( \mathrm{L}\right) $ in the left
area and $\bar{T}^{10}\left( \mathrm{R}\right) $ in the right area of the
axis $x$ are different due to the potential energy difference $\Delta U$.
These quantities correspond to the areas where the electric field is absent.
They are proportional to the factor $N_{n}^{\mathrm{cr}}$ formed due to an
electric field situated in a region{\large \ }$S_{\mathrm{int}}$. If the
electric field vanishes, $E\rightarrow 0$, in the area $S_{\mathrm{int}}$,
then at the same time the pair production vanishes, $N_{n}^{\mathrm{cr}%
}\rightarrow 0$. That is why the above densities characterize in a sense
real particles and why they cannot change after the electric field is
switched off.

The normal forms of the operators $J^{1}$\ and $T^{10}$\ with respect to the
\textrm{out}-vacuum are%
\begin{equation}
N_{\mathrm{out}}\left( J^{1}\right) =J^{1}-\left\langle 0,\mathrm{out}%
\left\vert J^{1}\right\vert 0,\mathrm{out}\right\rangle ,\;\;N_{\mathrm{out}%
}\left( T^{10}\right) =T^{10}-\left\langle 0,\mathrm{out}\left\vert
T^{10}\right\vert 0,\mathrm{out}\right\rangle .  \label{m9}
\end{equation}%
Then, taking into account Eqs. (\ref{m7}) and (\ref{m8}), we can calculate
the densities of the longitudinal current and energy flux corresponding to
the final particles as means in the initial vacuum state,%
\begin{eqnarray}
&&\tilde{J}_{\mathrm{cr}}^{1}=\left\langle N_{\mathrm{out}}\left(
J^{1}\right) \right\rangle _{\mathrm{in}}=\left\langle J^{1}(x)\right\rangle
_{\mathrm{in}}-\left\langle J^{1}(x)\right\rangle _{\mathrm{out}%
}=2\left\langle J^{1}(x)\right\rangle ^{p}=2\bar{J}^{1};  \notag \\
&&\left\langle \tilde{T}^{10}(x)\right\rangle _{\mathrm{cr}}=\left\langle N_{%
\mathrm{out}}\left[ T^{10}(x)\right] \right\rangle _{\mathrm{in}%
}=\left\langle T^{10}(x)\right\rangle _{\mathrm{in}}-\left\langle
T^{10}(x)\right\rangle _{\mathrm{out}}=2\left\langle T^{10}(x)\right\rangle
^{p}.  \label{m10}
\end{eqnarray}

The means (\ref{m10}) depend on the parameter $\tau ${\large . }In the range
$\Omega _{3}$\ this parameter can be interpreted as the time of observation
of the pair production and is related to the time{\large \ }$T${\large . }We
suppose that all the measurements are performed during a macroscopic time $T$
when the external field can be considered as constant. In this case, for
example, we believe that the longitudinal current of created particles is
equal to the flux density $N^{\mathrm{cr}}(TV_{\perp })^{-1}$ of the
particles times the charge $e$,{\large \ }%
\begin{equation}
J_{\mathrm{cr}}^{1}=eN^{\mathrm{cr}}(TV_{\perp })^{-1},  \label{m11}
\end{equation}%
where $N^{\mathrm{cr}}$\ is the total number of pairs\ created from the
vacuum during the time $T$,\ given by Eq. (\ref{TN})\footnote{%
Note that contributions to the sum (\ref{TN}) due to the tiny ranges of low
velocities, $v^{\mathrm{L/R}}\sim \left\vert x_{\mathrm{L/R}}\right\vert /T$%
, are negligible if the time{\large \ }$T$\ satisfies Eq. (\ref{m27}). In
the following we assume that these subranges are excluded from sums over the
range{\large \ }$\Omega _{3}$ and the condition (\ref{m0}) is satisfied for
any $v^{\mathrm{L/R}}$ under consideration.}{\large .} One can see that $%
\tilde{J}_{\mathrm{cr}}^{1}$ given by Eq. (\ref{m10} ) coincides with the
physically motivated expression (\ref{m11}) under the condition that the
times $\tau $\ and $T$\ coincide, i.e., $\tau =T${\large . }Such a relation
fixes the\ proposed renormalization procedure.

Thus, we have%
\begin{eqnarray}
&&J_{\mathrm{cr}}^{1}=2\left\langle J^{1}(x)\right\rangle ^{p}=e\sum_{n\in
\Omega _{3}}j_{n}^{1},\ \ j_{n}^{1}=N_{n}^{\mathrm{cr}}(TV_{\perp })^{-1};
\label{m11a} \\
&&\left\langle T^{10}(x)\right\rangle _{\mathrm{cr}}=2\left\langle
T^{10}(x)\right\rangle ^{p}=\left\{
\begin{tabular}{l}
$T_{\mathrm{cr}}^{10}\left( \mathrm{L}\right) =-\sum_{n\in \Omega _{3}}\pi
_{0}\left( \mathrm{L}\right) j_{n}^{1}\ ,\ \ x\in S_{\mathrm{L}}$ \\
$T_{\mathrm{cr}}^{10}\left( \mathrm{R}\right) =\sum_{n\in \Omega
_{3}}\left\vert \pi _{0}\left( \mathrm{R}\right) \right\vert j_{n}^{1}\ ,\ \
x\in S_{\mathrm{R}}$%
\end{tabular}%
\right. .  \label{m11b}
\end{eqnarray}%
Here $j_{n}^{1}$ is the flux density of\textrm{\ }particles created with
given $n$. The density $j_{n}^{1}$ and longitudinal current density $J_{%
\mathrm{cr}}^{1}$ are conserved in the $x$-direction. The density $j_{n}^{1}$
is formed by electrons on the left of the area $S_{\mathrm{int}}$ , whereas
it is formed by positrons on the right of the area $S_{\mathrm{int}}$.
Differential mean numbers of electrons and positrons from pairs with a given
$n$ are equal. The differential numbers $N_{n}^{\mathrm{cr}}$\ characterize
electrons created on the left of the area $S_{\mathrm{int}}$\ and at the
same time characterize positrons created on the right of the area{\large \ }$%
S_{\mathrm{int}}${\large . }Thus, on the left of the area $S_{\mathrm{int}}$
and on the right of the area $S_{\mathrm{int}}$ there are fluxes consisting
of all electrons and all positrons created from vacuum by the external
electric field, which, on the one hand, is natural due to physical
considerations, and on the other hand confirms the correctness of the
applied approach.

Using Eqs. (\ref{m3}), (\ref{m5.4}), and (\ref{m5.5}), we find that
averaging over the \textrm{in}- and \textrm{out}-vacuum of the charge and
diagonal components of EMT operators gives the same results,%
\begin{equation}
\left\langle J^{0}(x)\right\rangle _{\mathrm{in}}=\left\langle
J^{0}(x)\right\rangle _{\mathrm{out}},\ \ \left\langle T^{\mu \mu
}(x)\right\rangle _{\mathrm{in}}=\left\langle T^{\mu \mu }(x)\right\rangle _{%
\mathrm{out}}\ \ \mathrm{if}\ x\in S_{\mathrm{L}}\ \mathrm{or}\ S_{\mathrm{R}%
}.  \label{m12}
\end{equation}%
This coincidence (at least for a time) is an important distinguishing
feature of the \textrm{in} and \textrm{out} states in the theory under
consideration. In particular, the characteristics of the vacuum states do
not change over time $T$. However, the initial and final vacua differ due to
the corresponding fluxes of charges and energy through the surface $x=%
\mathrm{const}$.

One can see how quantities (\ref{m12}) are related to the direct
characteristics of the vacuum instability. To this end, in these quantities,
we separate the contributions of the matrix elements $\left\langle J^{\mu
}(x)\right\rangle ^{c}$ and $\left\langle T_{\mu \nu }(x)\right\rangle ^{c}$
given in Eq. (\ref{m5.3b}) on the left of the area $S_{\mathrm{int}}$ and on
the right of the area $S_{\mathrm{int}}$. Using the representations (\ref%
{m5.4}), one can see that $\tilde{S}^{-}(X,X^{\prime })$ on the area $x\in
S_{\mathrm{R}}$ and $\tilde{S}^{+}(X,X^{\prime })$ on the area $x\in S_{%
\mathrm{L}}$ represent $x$-dependent interference terms, containing factors $%
\sim \exp \left[ i2\left\vert p^{\mathrm{R}}\right\vert \left( x-x_{\mathrm{R%
}}\right) \right] $ and $\exp \left[ -i2\left\vert p^{\mathrm{L}}\right\vert
\left( x-x_{\mathrm{L}}\right) \right] $ respectively. Such terms do not
contribute to the fluxes $\left\langle J^{1}(x)\right\rangle ^{c}$ and $%
\left\langle T^{10}(x)\right\rangle ^{c}$. They do not contribute to any
integrals over $x$ and, therefore, they can be neglected in the densities $%
\left\langle J^{0}(x)\right\rangle ^{c}$ and $\left\langle T^{\mu \mu
}(x)\right\rangle ^{c}$. Now we analyze the opposite case, namely, let us
consider representations of the function $\tilde{S}^{-}(X,X^{\prime })$ in
the area $x\in S_{\mathrm{L}}$ and of the function $\tilde{S}%
^{+}(X,X^{\prime })$ on the area $x\in S_{\mathrm{R}}$. Using Eqs. (\ref%
{rel1}) and (\ref{m14}), and ignoring the $x$-dependent interference terms,
we obtain%
\begin{eqnarray}
&&\tilde{S}^{-}(X,X^{\prime })=\frac{i}{2}\sum_{n\in \Omega _{3}}\left[
_{+}\psi _{n}\left( X\right) \;_{+}\bar{\psi}_{n}\left( X^{\prime }\right)
+\;_{-}\psi _{n}\left( X\right) \;_{-}\bar{\psi}_{n}\left( X^{\prime
}\right) \right] ,\;x,x^{\prime }\in S_{\mathrm{L}}\ ,  \notag \\
&&\tilde{S}^{+}(X,X^{\prime })=\frac{i}{2}\sum_{n\in \Omega _{3}}\left[
^{+}\psi _{n}\left( X\right) \;^{+}\bar{\psi}_{n}\left( X^{\prime }\right)
+\;^{-}\psi _{n}\left( X\right) \;^{-}\bar{\psi}_{n}\left( X^{\prime
}\right) \right] ,\;x,x^{\prime }\in S_{\mathrm{R}}.  \label{m15}
\end{eqnarray}%
Since singular functions (\ref{m15}) are constructed with the help of
solutions of the Dirac equation for free particles (in the regions $S_{%
\mathrm{L}}$ and $S_{\mathrm{R}}$, respectively), they do not contribute to
the fluxes under consideration. Thus,%
\begin{equation}
\left\langle J^{1}(x)\right\rangle ^{c}=0,\;\left\langle
T^{10}(x)\right\rangle ^{c}=0,\ \ x\in S_{\mathrm{L}}\;\mathrm{or}\;S_{%
\mathrm{R}},  \label{m15b}
\end{equation}%
which means that the fluxes calculated with the help of $S_{\text{\textrm{in}%
}}^{c}$ and $S^{p}$ coincide.

The quantities $\left\langle J^{0}(x)\right\rangle _{\mathrm{in}}\ $and$\
\left\langle T^{\mu \mu }(x)\right\rangle _{\mathrm{in}}${\large \ }at%
{\large \ }$x\in S_{\mathrm{L/R}}$\ have to be renormalized. The procedure
below has not been considered previously in QED with $x$-electric steps.%
{\large \ }Let us discuss contributions to the quantities $\left\langle
J^{0}(x)\right\rangle ^{c}$\ and $\left\langle T^{\mu \mu }(x)\right\rangle
^{c}$ that are stipulated by the singular functions (\ref{m15}) at the
regions $S_{\mathrm{L}}$\ and $S_{\mathrm{R}}$.{\large \ }First we note that
the definition of the range $\Omega _{3}$, given by Eq. (\ref{ranges}), does
not depend on variations of the electric field in the region $S_{\mathrm{int}%
}$ which leave unchanged the potentials $U_{\mathrm{L}}$\ and $U_{\mathrm{R}%
} $ -- and therefore neither do the singular functions under consideration.%
{\large \ }These contributions from the left and from the right are
different; this difference is due to different kinetic energies $\pi
_{0}\left( \mathrm{L/R}\right) $ on the left of the area $S_{\mathrm{int}}$
and on the right of the area $S_{\mathrm{int}}$. The quantities $\pi
_{0}\left( \mathrm{L/R}\right) $ are defined by potentials $U_{\mathrm{L}}$
and $U_{\mathrm{R}}$, respectively. The difference of these potentials gives
us the magnitude of the $x$-step, $\pi _{0}\left( \mathrm{L}\right) -\pi
_{0}\left( \mathrm{R}\right) =U_{\mathrm{R}}-U_{\mathrm{L}}=\Delta U$, which
defines the Klein range $\Omega _{3}$. It is clear that the difference $%
\Delta U$ is produced by work done by the electric field in the area $S_{%
\mathrm{int}}$. However, it should be borne in mind that this work can be
done by an electric field of any form and any intensity, including an
arbitrarily low intensity, if the area $S_{\mathrm{int}}$ is wide enough.
Exactly in such a way the inhomogeneous structure of the vacuum in the
presence of an $x$-step out of the field area $S_{\mathrm{int}}$ is
manifested. This effect is similar to a vacuum effect in the space around a
magnetic flux confined inside a solenoid, i.e., in the presence of an
Aharonov-Bohm potential. The mean numbers of pairs created from the vacuum, $%
N_{n}^{\mathrm{cr}}$, are not determined by the magnitude $\Delta U$, but by
the electric field intensity $E\left( x\right) $, so that for small fields, $%
E\left( x\right) \rightarrow 0$, these quantities are negligible, $N_{n}^{%
\mathrm{cr}}\rightarrow 0$. Thus, nonzero values of the densities $%
\left\langle J^{0}(x)\right\rangle ^{c}$ and $\left\langle T^{\mu \mu
}(x)\right\rangle ^{c}$ out of the area $S_{\mathrm{int}}$ have nothing to
do with real particles and with the vacuum polarization due to the fact that
the electric field is absent there. These terms represent unobservable
quasilocal contributions to the vacuum of free particles and they should be
neglected in the spirit of the idea of renormalization.

Thus, we believe that the means $\left\langle J^{0}(x)\right\rangle _{%
\mathrm{in}}\ $and$\ \left\langle T^{\mu \mu }(x)\right\rangle _{\mathrm{in}%
} $ at $x\in S_{\mathrm{L/R}}\;$have to be renormalized to the ones $%
\left\langle J^{0}(x)\right\rangle _{\mathrm{in}}^{\mathrm{ren}}$ and $%
\left\langle T^{\mu \mu }(x)\right\rangle _{\mathrm{in}}^{\mathrm{ren}}$ as
follows:%
\begin{eqnarray}
&&\left\langle J^{0}(x)\right\rangle _{\mathrm{in}}^{\mathrm{ren}%
}=\left\langle J^{0}(x)\right\rangle _{\mathrm{in}}-\left\langle
J^{0}(x)\right\rangle ^{c}=\left\langle J^{0}(x)\right\rangle ^{p},  \notag
\\
&&\left\langle T^{\mu \mu }(x)\right\rangle _{\mathrm{in}}^{\mathrm{ren}%
}=\left\langle T^{\mu \mu }(x)\right\rangle _{\mathrm{in}}-\left\langle
T^{\mu \mu }(x)\right\rangle ^{c}=\left\langle T^{\mu \mu }(x)\right\rangle
^{p}\ .  \label{m16}
\end{eqnarray}%
Note that fluxes (\ref{m7}) and (\ref{m8}) can be represented in the same
manner with account taken of Eqs. (\ref{m15b}). Using Eqs. (\ref{rel1}), we
find%
\begin{eqnarray}
&&\left\langle J^{0}(x)\right\rangle ^{p}=\left\{
\begin{array}{c}
-\ \bar{J}^{0}\left( \mathrm{L}\right) ,\ \ x\in S_{\mathrm{L}} \\
\bar{J}^{0}\left( \mathrm{R}\right) ,\ \ x\in S_{\mathrm{R}}%
\end{array}%
\right. ,  \notag \\
&&\bar{J}^{0}\left( \mathrm{L/R}\right) =\frac{e}{2}\sum_{n\in \Omega
_{3}}j_{n}^{0}\left( \mathrm{L/R}\right) ,\ \ j_{n}^{0}\left( \mathrm{L/R}%
\right) =j_{n}^{1}/v^{\mathrm{L/R}},  \label{m17}
\end{eqnarray}%
where $j_{n}^{1}$ is given by Eq. (\ref{m11a}), and%
\begin{eqnarray}
&&\left\langle T^{00}(x)\right\rangle ^{p}=\left\{
\begin{array}{c}
\frac{1}{2}\sum_{n\in \Omega _{3}}j_{n}^{0}\left( \mathrm{L}\right) \pi
_{0}\left( \mathrm{L}\right) ,\ \ x\in S_{\mathrm{L}} \\
\frac{1}{2}\sum_{n\in \Omega _{3}}j_{n}^{0}\left( \mathrm{R}\right)
\left\vert \pi _{0}\left( \mathrm{R}\right) \right\vert ,\ \ x\in S_{\mathrm{%
R}}%
\end{array}%
\right. ,  \notag \\
&&\left\langle T^{11}(x)\right\rangle ^{p}=\left\{
\begin{array}{c}
\frac{1}{2}\sum_{n\in \Omega _{3}}j_{n}^{1}\left\vert p^{\mathrm{L}%
}\right\vert ,\mathrm{\ }\ x\in S_{\mathrm{L}} \\
\frac{1}{2}\sum_{n\in \Omega _{3}}j_{n}^{1}\left\vert p^{\mathrm{R}%
}\right\vert ,\ \ x\in S_{\mathrm{R}}%
\end{array}%
\right. ,\   \notag \\
&&\left\langle T^{kk}(x)\right\rangle ^{p}=\left\{
\begin{array}{c}
\frac{1}{2}\sum_{n\in \Omega _{3}}j_{n}^{1}\left( p_{k}\right)
^{2}/\left\vert p^{\mathrm{L}}\right\vert ,\ \ x\in S_{\mathrm{L}} \\
\frac{1}{2}\sum_{n\in \Omega _{3}}j_{n}^{1}\left( p_{k}\right)
^{2}/\left\vert p^{\mathrm{R}}\right\vert ,\ \ x\in S_{\mathrm{R}}%
\end{array}%
\right. ,\ k\neq 1.  \label{m18}
\end{eqnarray}

Next, we discuss how it is possible to extract physically meaningful
information from the obtained formal expressions{\large \ }(\ref{m17}) and (%
\ref{m18}){\large . }Let us calculate the charge density and the
longitudinal current density of the particles created. To this end we turn
to Eq. (\ref{m11a}). We see that the created electrons with a given $n$ are
moving to the left with the velocity $v^{\mathrm{L}}$, whereas the current
density $ej_{n}^{1}$ is moving to the right. During the time $T$ these
electrons carry the charge $ej_{n}^{1}T$ over a unit area $V_{\bot }$ of the
surface $x=x_{\mathrm{L}}$. Taking into account that this charge is evenly
distributed over the cylindrical volume of the length $v^{\mathrm{L}}T$, we
see that the charge density of the created electrons with given $n$ is $%
ej_{n}^{1}/\left( -v^{\mathrm{L}}\right) $. Note that this quantity
coincides with the density $-ej_{n}^{0}\left( \mathrm{L}\right) $, where $%
j_{n}^{0}\left( \mathrm{L}\right) $ is given by Eq. (\ref{m17}), $%
-ej_{n}^{0}\left( \mathrm{L}\right) =ej_{n}^{1}/\left( -v^{\mathrm{L}%
}\right) $. The created positrons with a given $n$ are moving to the right
with the velocity $v^{\mathrm{R}}$, the current density $ej_{n}^{1}$ moves
in the same direction. During the time $T$ these positrons carry the same
charge $ej_{n}^{1}T$ over a unit area $V_{\bot }$ of the surface $x=x_{%
\mathrm{R}}$. This charge is evenly distributed over the cylindrical volume
of the length $v^{\mathrm{R}}T$. Thus, the charge density of the created
positrons with a given $n$ is $ej_{n}^{1}/v^{\mathrm{R}}$. This quantity
coincides with the density $ej_{n}^{0}\left( \mathrm{R}\right) $, where $%
j_{n}^{0}\left( \mathrm{R}\right) $ is given by Eq. (\ref{m17}), $%
ej_{n}^{0}\left( \mathrm{R}\right) =ej_{n}^{1}/v^{\mathrm{R}}$. This means
that the charge density of created pairs is%
\begin{equation}
J_{\mathrm{cr}}^{0}(x)=\left\{
\begin{array}{c}
-e\sum_{n\in \Omega _{3}}j_{n}^{0}\left( \mathrm{L}\right) ,\mathrm{\ }\
x\in S_{\mathrm{L}} \\
e\sum_{n\in \Omega _{3}}j_{n}^{0}\left( \mathrm{R}\right) ,\mathrm{\ }\ x\in
S_{\mathrm{R}}%
\end{array}%
\right. .\   \label{m19}
\end{equation}%
Comparing Eqs. (\ref{m17}) and (\ref{m19})$\ $we can relate the quantity%
{\large \ }$J_{\mathrm{cr}}^{0}(x)${\large \ }with the one{\large \ }$%
\left\langle J^{0}(x)\right\rangle ^{p}$. One can easily see that $J_{%
\mathrm{cr}}^{0}(x)$ and $J_{\mathrm{cr}}^{1}$ are two nonzero components of
the $d$ dimensional Lorentz vector $J_{\mathrm{cr}}^{\mu }(x)$ (note that
its transversal components are zero) that represents the current density of
the created pairs,
\begin{eqnarray}
&&J_{\mathrm{cr}}^{0}(x)=2\left\langle J^{0}(x)\right\rangle ^{p},\;J_{%
\mathrm{cr}}^{k}(x)=2\left\langle J^{k}(x)\right\rangle ^{p}=0\;\mathrm{if}%
\;k\neq 1;  \notag \\
&&J_{\mathrm{cr}}^{1}(x)=2\left\langle J^{1}(x)\right\rangle ^{p},\mathrm{\ }%
\ x\in S_{\mathrm{L}}\;\mathrm{or}\;S_{\mathrm{R}}.  \label{m20}
\end{eqnarray}%
In the same manner, one can derive the following relations for the nonzero
components of EMT:%
\begin{equation}
T_{\mathrm{cr}}^{\mu \mu }(x)=2\left\langle T^{\mu \mu }(x)\right\rangle
^{p};\ \ T_{\mathrm{cr}}^{10}(x)=2\left\langle T^{10}(x)\right\rangle ^{p},%
\mathrm{\ }\ x\in S_{\mathrm{L}}\;\mathrm{or}\;S_{\mathrm{R}}.  \label{m21}
\end{equation}%
Thus, Eqs. (\ref{m20}) and (\ref{m21}) give the opportunity derive all the
characteristics of created particles with the help of the function $S^{p}$
given by Eq. (\ref{m13.2}). Note that the physical quantities related to the
particles created in the areas $S_{\mathrm{L}}$ and $S_{\mathrm{R}}$ satisfy
the relations
\begin{eqnarray}
&&J_{\mathrm{cr}}^{0}(x_{\mathrm{L}})=-J_{\mathrm{cr}}^{0}(x_{\mathrm{R}}),\
\ J_{\mathrm{cr}}^{1}(x_{\mathrm{L}})=J_{\mathrm{cr}}^{1}(x_{\mathrm{R}}),
\notag \\
&&T_{\mathrm{cr}}^{\mu \mu }(x_{\mathrm{L}})=T_{\mathrm{cr}}^{\mu \mu }(x_{%
\mathrm{R}}),\ \ T_{\mathrm{cr}}^{10}(x_{\mathrm{L}})=-T_{\mathrm{cr}%
}^{10}(x_{\mathrm{R}}).  \label{m22}
\end{eqnarray}

Thus, in QED with critical $x$-step, there exist relations between the
nonzero physical quantities (\ref{m20}), (\ref{m21}) and the \textrm{in--in}
(\textrm{out--out}) mean values (\ref{m12}) and (\ref{m16}),
\begin{eqnarray}
&&J_{\mathrm{cr}}^{0}(x)=\left\langle J^{0}(x)\right\rangle _{\mathrm{in}}^{%
\mathrm{ren}}+\left\langle J^{0}(x)\right\rangle _{\mathrm{out}}^{\mathrm{ren%
}},  \notag \\
&&J_{\mathrm{cr}}^{1}(x)=\left\langle J^{1}(x)\right\rangle _{\mathrm{in}%
}-\left\langle J^{1}(x)\right\rangle _{\mathrm{out}}\;;  \notag \\
&&T_{\mathrm{cr}}^{\mu \mu }(x)=\left\langle T^{\mu \mu }(x)\right\rangle _{%
\mathrm{in}}^{\mathrm{ren}}+\left\langle T^{\mu \mu }(x)\right\rangle _{%
\mathrm{out}}^{\mathrm{ren}},  \notag \\
&&T_{\mathrm{cr}}^{10}(x)=\left\langle T^{10}(x)\right\rangle _{\mathrm{in}%
}-\left\langle T^{10}(x)\right\rangle _{\mathrm{out}},\ x\in S_{\mathrm{L}}\;%
\mathrm{or}\;S_{\mathrm{R}}.  \label{m23}
\end{eqnarray}%
We see, for example, that $\left\langle J^{0}(x)\right\rangle _{\mathrm{in}%
}^{\mathrm{ren}}$ and $\left\langle J^{1}(x)\right\rangle _{\mathrm{in}}$
are two nonzero components of the $d$ dimensional Lorentz vector $%
\left\langle J^{\mu }(x)\right\rangle _{\mathrm{in}}^{\mathrm{ren}%
}=\left\langle J^{\mu }(x)\right\rangle ^{p},$ whereas $\left\langle
J^{0}(x)\right\rangle _{\mathrm{out}}^{\mathrm{ren}}$ and $-\ \left\langle
J^{1}(x)\right\rangle _{\mathrm{out}}$ are two nonzero components of the
same Lorentz vector. A similar relation holds for the vacuum means of EMT.
We stress that the calculated physical quantities (\ref{m23}) are determined
by precisely the final particles.{\large \ }Thus, the proposed procedure
allows one to relate the characteristics of the created particles with
quasilocal quantities.

\section{Backreaction and consistency conditions\label{S4}}

The regularization and renormalization procedures discussed above allow you
to expand the range of tasks to which  QED with $x$-steps concentrated in
restricted space areas can be applied.{\large \ }In such a way we study
below the effects of the backreaction on the vacuum instability, which
allows one, in turn, to establish the so-called consistency conditions. In
high-energy physics problems it is usually assumed that just from the
beginning there exists a classical electric field having a given energy. The
system of fermions interacting with this field is closed, that is, the total
energy of the system is conserved. In the  following we work under such an
assumption.\footnote{%
One can, however, imagine an alternative situation when these charges are
getting out of the regions $S_{\mathrm{L}}$\ and $S_{\mathrm{R}}$\ with the
help of the work done by an external storage battery. For example, dealing
with graphene devices, it is natural to assume that the constant strength on
the graphene plane is due to the applied fixed voltage, i.e., we are dealing
with an open system of fermions interacting with a classical electromagnetic
field. In that case there would be no backreaction problem. Note that the
evolution of the mean electromagnetic field in the graphene, taking into
account the backreaction of the matter field to the applied time-dependent
external field, was considered in Ref. \cite{GavGitY12}.}

Let us consider a volume $V=V_{\bot }\left( x_{\mathrm{R}}-x_{\mathrm{L}%
}\right) ,$ which contains the area $S_{\mathrm{int}}=\left( x_{\mathrm{L}%
},x_{\mathrm{R}}\right) .$ The total energy of created particles in the
volume $V$ is given by the corresponding volume integral of the energy
density $T_{\mathrm{cr}}^{00}\left( t,x\right) .$ The corresponding energy
conservation reads%
\begin{equation}
\frac{\partial }{\partial t}\int_{V_{\bot }}d\mathbf{r}_{\bot }\int_{x_{%
\mathrm{L}}}^{x_{\mathrm{R}}}T_{\mathrm{cr}}^{00}\left( t,x\right)
dx=-\oint_{\Sigma }T_{\mathrm{cr}}^{k0}(x)df_{k},  \label{m24}
\end{equation}%
where $\Sigma $ is a $d-1$ dimensional surface surrounding the volume $V$
and $df_{k}$, $k=1,...,d-1$, are the components of the surface element $d%
\mathbf{f}$. Taking into account that $T_{\mathrm{cr}}^{00}\left( t,x\right)
$ in Eq. (\ref{m24}) does not depend on transversal coordinates, $T_{\mathrm{%
cr}}^{k0}(x)=0$ for $k\neq 1$ and using Eqs. (\ref{m11a}) and (\ref{m11b})
we finf that the rate of the energy density change of the created particles
in the area $S_{\mathrm{int}}$ per unit of the spatial volume of the
hypersurface orthogonal to the $x$-direction is:
\begin{equation}
\frac{\partial }{\partial t}\int_{x_{\mathrm{L}}}^{x_{\mathrm{R}}}T_{\mathrm{%
cr}}^{00}\left( t,x\right) dx=T_{\mathrm{cr}}^{10}\left( \mathrm{L}\right)
-T_{\mathrm{cr}}^{10}\left( \mathrm{R}\right) =-\Delta U\frac{N^{\mathrm{cr}}%
}{TV_{\perp }},  \label{m25}
\end{equation}%
where $N^{\mathrm{cr}}$, given\ by Eq. (\ref{TN}), is the total number of
pairs\ created from the vacuum. It characterizes the loss of the energy that
created particles carry away from the region $S_{\mathrm{int}}$. At the same
time, the constant rate (\ref{m25}) determines the power of the constant
electric field spent on the pair creation. Integrating this rate over the
time duration of an electric field from $t_{\mathrm{in}}$ to $t_{\mathrm{out}%
}$, and using the notation%
\begin{equation*}
\Delta T_{\mathrm{cr}}^{00}\left( x\right) =-\int_{t_{\mathrm{in}}}^{t_{%
\mathrm{out}}}\frac{\partial }{\partial t}T_{\mathrm{cr}}^{00}\left(
t,x\right) dt\ ,
\end{equation*}%
we find the total energy density of created pairs per unit of the orthogonal
hypersurface as%
\begin{equation}
\int_{x_{\mathrm{L}}}^{x_{\mathrm{R}}}\Delta T_{\mathrm{cr}}^{00}\left(
x\right) dx=\Delta U\frac{N^{\mathrm{cr}}}{V_{\perp }}.  \label{m26}
\end{equation}

We assume that the constant external electric field which represents the $x$%
-step exists during a macroscopically large time period $T$ satisfying the
condition (\ref{m27}).

Now we are able to discuss the backreaction problem.\textrm{\ }It is clear
that while creating pairs from the vacuum the external electric field is
losing its energy and should change (weaken) with time.\textrm{\ }Thus, the
applicability of the constant field approximation, which is used in the
formulation of $QED$ with $x$-step, is limited by the smallness of the
backreaction. The results obtained above allow us to find conditions that
provide this smallness, we call these consistency conditions in what
follows. These conditions can be obtained from the requirement that the
energy density (\ref{m26}) is essentially smaller than the energy density of
the external electric field (per unit of the orthogonal hypersurface).

To do this, we consider the case of an uniform electric field confined
between two capacitor plates separated by a finite distance $L$. We call
such a configuration an $L$-constant electric field. In this case $\Delta
U=eEL$. The particle creation in such a case was studied in Ref. \cite%
{L-field}. If%
\begin{equation}
L\gg \left( eE\right) ^{-1/2}\max \left\{ 1,m^{2}/eE\right\} ,  \label{m28}
\end{equation}%
the $L$-constant electric field can be expected to simulate a small-gradient
electric field. Besides, in such case boundary effects are negligible and
the density $\Delta T_{\mathrm{cr}}^{00}\left( x\right) $ is uniform in the
leading term approximation, $\Delta T_{\mathrm{cr}}^{00}\left( x\right)
=\Delta T_{\mathrm{cr}}^{00}$. In this case\ the number density in the RHS
of Eq. (\ref{m26}) has the form%
\begin{equation}
\frac{N^{\mathrm{cr}}}{V_{\perp }}=Ln^{\mathrm{cr}},\;n^{\mathrm{cr}}=Tr^{%
\mathrm{cr}},\;r^{\mathrm{cr}}=\frac{J\left( eE\right) ^{d/2}}{(2\pi )^{d-1}}%
\exp \left( -\pi \frac{m^{2}}{eE}\right) ,  \label{m29}
\end{equation}%
where $n^{\mathrm{cr}}$ is the number density of pairs created per unit
volume of the field area and $J=2^{[d/2]-1}$ is the number of spin degrees
of freedom ($J=1$ for scalar particles) \cite{L-field}. Taking into account
that $x_{\mathrm{R}}-x_{\mathrm{L}}=L$ and integrating over $x$ in Eq. (\ref%
{m26}) we find that the total energy density of created pairs per the volume
unit is
\begin{equation}
\Delta T_{\mathrm{cr}}^{00}=eELTr^{\mathrm{cr}}.  \label{m30}
\end{equation}%
Note that the density $r^{\mathrm{cr}}$, given by Eq. (\ref{m29}), can be
identified with the pair-production rate in the case of a constant uniform
external electric field; see Ref.~\cite{GavG96a}. The number density $n^{%
\mathrm{cr}}$, given by Eq. (\ref{m29}), coincides with result obtained for
the case of the $T$-constant field (a uniform electric field which
effectively acts during a sufficiently large but finite time interval $T$)
\cite{GavG96a}.

Note that unlike the case of the pair creation by the $T$-constant field,
considered in the framework of QED with $t$-step, where $N^{\mathrm{cr}}$ is
a linear function of the finite time duration of the field $T$ and $%
L\rightarrow \infty $ is a given constant, $N^{\mathrm{cr}}$ given by Eq.~(%
\ref{m29}) is a linear function of the field length $L$ and $T\rightarrow
\infty $ is a given constant. The $T$-constant and $L$-constant fields are
physically distinct. Note that while in the $T$-constant field model it is
allowed that $L/T\sim 1$ in the case under consideration (of a finite field
area) we assumed that Eq. (\ref{m0}) is satisfied. The latter assumption
implies the inequality $T\gg L$. If we use the fact that the pair-production
rate $r^{\mathrm{cr}}$, given by Eq. (\ref{m29}), is the same both for $T$%
-constant and $L$-constant fields, then we can compare energy densities of
created pairs per unit volume for both cases. The energy density of the
created pairs $\langle T_{00}(t_{\mathrm{out}})\rangle $ per unit volume for
the case of $T$-constant field was derived in Refs. \cite{GG06-08,GavGitY12}
and it has the form%
\begin{equation}
\langle T_{00}(t_{\mathrm{out}})\rangle =eET^{2}r^{\mathrm{cr}}.  \label{m31}
\end{equation}%
The densities (\ref{m30}) and (\ref{m31}) are related as follows:%
\begin{equation}
\Delta T_{\mathrm{cr}}^{00}=\frac{L}{T}\langle T_{00}(t_{\mathrm{out}%
})\rangle ,  \label{m32}
\end{equation}%
which means that $\Delta T_{\mathrm{cr}}^{00}\ll \langle T_{00}(t_{\mathrm{%
out}})\rangle $. Thus, taking into account that the area occupied by the
external field is finite significantly reduces the energy densities of the
created pairs per unit volume for given field strength $E$ and its duration $%
T$. In turn, this also affects the degree of the backreaction.

In problems of high-energy physics it is usually assumed that just from the
beginning there exists a classical electric field having a given energy
density. The system of fermions interacting with this field is supposed to
be closed, that is, its total energy is conserved. Such an assumption is
consistent only if the backreaction due to the pair creation is relatively
small with respect to the background. For the same reasons QED with either
strong $T$-constant or $L$-constant fields can be considered as a consistent
model only if the backreaction is small. Below we derive the conditions
which provide such a smallness for QED with $x$-step. Similar to the case of
QED with $t$-step (see Ref. \cite{GavG08}), we call them  consistency
conditions.

The consistency conditions follow from the supposition that the energy
density $\Delta T_{\mathrm{cr}}^{00}$, arising precisely due to the action
of a $L$-constant electric field, should be essentially smaller than the
energy density of the external electric field. As such a density in $d=4$
dimensions we take the classical energy density $\mathcal{E}^{\left(
0\right) }$ of the electric field, $\mathcal{E}^{\left( 0\right)
}=E^{2}/8\pi $. Thus, the condition of the smallness of the backreaction can
be written as $\Delta T_{\mathrm{cr}}^{00}\ll E^{2}/8\pi $. Taking into
account Eq. (\ref{m30}), we obtain the sought-for consistency conditions as
a restriction from above on the dimensionless parameter $eELT$:%
\begin{equation}
eELT\ll \frac{\pi ^{2}}{J\alpha }\,\exp \left( \pi \frac{m^{2}}{eE}\right) .
\label{m33}
\end{equation}%
Here $\alpha $ is the fine structure constant, $\alpha =e^{2}/c\hslash =e^{2}
$, and $J=2$.

Note that using the appropriate number of the spin degrees of freedom
inequality (\ref{m33}) can be generalized for scalar particles ($J=1$) and
for vector particles ($J=3$). On the other hand, all the asymptotic formulas
have been obtained under conditions (\ref{m27}) and (\ref{m28}), which
impose restrictions on the parameter $eELT$ from below,
\begin{equation}
eELT\gg \max \left\{ 1,\left( m^{2}/eE\right) ^{2}\right\} \ .  \label{m34}
\end{equation}%
One can easily extend these results to $d$\ dimensions, using the
expressions for $r^{\mathrm{cr}}$, given by Eq. (\ref{m29}).

Since $\pi ^{2}/J\alpha \gg 1$, there exists a range of values of parameters
$E$, $L$ and $T$ that satisfies both the inequalities. We recall once again
that the consistency conditions in the case of QED with $t$-step
(specifically for the case of the $T$-constant field) were obtained in Ref.
\cite{GG06-08}. They have the following form%
\begin{equation}
\max \left\{ 1,\left( m^{2}/eE\right) ^{2}\right\} \ll eET^{2}\ll \frac{\pi
^{2}}{J\alpha }\,\exp \left( \pi \frac{m^{2}}{eE}\right) .  \label{m35}
\end{equation}

Taking into account that in the case under consideration the dimensionless
parameter satisfies the inequality $eELT\ll eET^{2}$, one can see that the
consistency condition (\ref{m33}) is much less restrictive from above than
the one of  (\ref{m35}).

\section{Brief summary\label{S7}}

We consider the present article as a natural and important addition to the
nonperturbative formulation of QED with $x$-steps presented in Ref. \cite%
{x-case}. There, we have calculated global effects (global quantities) of
zeroth order with respect to the radiative interaction, as regards the
vacuum-to-vacuum transition amplitude, mean differential and total numbers
of created particles, and so on. Here we propose a new renormalization and
volume regularization{\large \ }procedures which allow one to calculate and
distinguish physical parts of different matrix elements of the operators of
the current and of the energy-momentum tensor, at the same time relating the
latter with characteristics of the vacuum instability.{\large \ }The
renormalization and volume regularization procedures are{\large \ }%
associated with the introduction of a modified inner product and a parameter
$\tau $ of the regularization. Based on physical considerations, we fix this
parameter. It turns out that in the Klein range this parameter can be
interpreted as the  time of the observation of the pair production process.
We derive conditions that provide the time independence of the introduced
inner product.

With the help of the calculated characteristics of the vacuum instability,
we consider the problem of the backreaction problem in QED with $x$-steps.
In the case of an uniform electric field confined between two capacitor
plates separated by a finite distance $L$, we see that the smallness of the
backreaction implies a restriction on the dimensionless parameter $eELT$. We
see this consistency condition is much less restrictive from above than the
one derived for QED with $t$-steps in our work \cite{GavG08}.

It should be noted that, recently, there was proposed a new formulation of
locally constant field approximations (LCFAs) in QED with{\large \ }$x$%
-steps, which does not rely on the Heisenberg--Euler action \cite%
{GavGitSh19D} (similar approximation was formulated in QED with $t$-steps
slowly varying with time \cite{GavGit17}).{\large \ }As part of this
formulation,\ we have constructed universal approximate representations for
the total number and current density of the created particles in arbitrary
weakly inhomogeneous $x$-steps.{\large \ }We hope that the regularization
and renormalization procedures proposed in the present article will allow us
to formulate soon a LCFA adequate for calculating the vacuum means of the
current density and EMT in QED with arbitrary weakly inhomogeneous $x$-steps.

\subparagraph{\protect\large Acknowledgement}

The work of both authors is supported by Russian Science Foundation (Grant
no. 19-12-00042).

\appendix

\section{Some properties of solutions of the Dirac equation with critical $x$%
-steps \label{Ap}}

Here we briefly recall some features of the solutions of the Dirac equation
with critical $x$-steps established in Ref. \cite{x-case} (see Sect. III and
Appendix B).

The Dirac equation with a $x$-steps has the form Eq. (\ref{ap.2a})%
\begin{equation}
i\partial _{0}\psi (X)=\hat{H}\psi (X),\ \ \hat{H}=\gamma ^{0}\left(
-i\gamma ^{j}\partial _{j}+m\right) +U(x),\ \ j=1,\ldots D,  \label{ap.2a}
\end{equation}%
and its solutions are of the form
\begin{align}
& \psi (X)=\exp \left( -ip_{0}t+i\mathbf{p}_{\perp }\mathbf{r}_{\perp
}\right) \psi _{n}(x),\ \mathbf{p}_{\perp }=\left( p^{2},\ldots
,p^{D}\right) ,  \notag \\
& \psi _{n}(x)=\left\{ \gamma ^{0}\left[ p_{0}-U(x)\right] -\gamma ^{1}\hat{p%
}_{1}-\mathbf{\gamma }_{\perp }\mathbf{p}_{\perp }+m\right\} \varphi
_{n}^{(\chi )}(x)\upsilon _{\chi },\   \notag \\
& \gamma ^{0}\gamma ^{1}\upsilon _{\chi }=\chi \upsilon _{\chi },\ \ \chi
=\pm 1,\ \ \upsilon _{\chi ^{\prime },\sigma ^{\prime }}^{\dag }\upsilon
_{\chi ,\sigma }=\delta _{\chi ^{\prime }\chi }\delta _{\sigma ^{\prime
}\sigma }.  \label{ap.2}
\end{align}%
Here $\psi (x)$ is a $2^{[d/2]}$-component spinor, $[d/2]$ stands for the
integer part of $d/2$, $\gamma ^{\mu }$ are the $\gamma $-matrices in $d$
dimensions, and $v_{\chi ,\sigma }$,{\large \ }$\sigma =(\sigma _{1},\sigma
_{2},\dots ,\sigma _{\lbrack d/2]-1})$,{\large \ }$\sigma _{s}=\pm 1$, is a
set of constant orthonormalized spinors. Scalar functions $\varphi
_{n}^{(\chi )}(x)$\ satisfy the following second-order differential equation:%
\begin{equation}
\left\{ \hat{p}_{x}^{2}-i\chi U^{\prime }\left( x\right) -\left[
p_{0}-U\left( x\right) \right] ^{2}+\mathbf{p}_{\bot }^{2}+m^{2}\right\}
\varphi _{n}^{(\chi )}(x)=0.  \label{e3}
\end{equation}%
These solutions are parametrized by the set of quantum numbers $n=(p_{0},%
\mathbf{p}_{\bot },\sigma )$. To construct complete sets necessary for our
purposes it is sufficient to choose only one value of $\chi .$ Furthermore,
for convenience, we choose a specific fixation, and then do not use this
index at all, $\varphi _{n}^{(\chi )}(x)=\varphi _{n}\left( x\right) $.

We construct two types of complete sets of solution in the form (\ref{ap.2}%
). The first one $_{\;\zeta }\psi _{n}\left( X\right) $ is defined by the
functions $\varphi _{n}\left( x\right) $ denoted as $_{\;\zeta }\varphi
_{n}\left( x\right) $ and the second one,$^{\;\zeta }\psi _{n}\left(
X\right) $, is defined by functions $\varphi _{n}\left( x\right) $ denoted $%
^{\;\zeta }\varphi _{n}\left( x\right) $. Asymptotically, these functions
have the forms
\begin{eqnarray}
&&_{\;\zeta }\varphi _{n}\left( X\right) =\ _{\zeta }\mathcal{N}\exp \left[
ip^{\mathrm{L}}\left( x-x_{\mathrm{L}}\right) \right] ,\ x\in S_{\mathrm{L}},
\notag \\
&&^{\;\zeta }\varphi _{n}\left( X\right) =\ ^{\zeta }\mathcal{N}\exp \left[
ip^{\mathrm{R}}\left( x-x_{\mathrm{R}}\right) \right] ,\ x\in S_{\mathrm{R}},
\label{m3}
\end{eqnarray}%
where $p^{\mathrm{L}}$ and $p^{\mathrm{R}}$ are given by Eq. (\ref{m3a}).
Thus, asymptotically the solutions $_{\;\zeta }\psi _{n}\left( X\right) $
and $^{\;\zeta }\psi _{n}\left( X\right) $ describe particles with given
real momenta $p^{\mathrm{L/R}}$ along the $x$ axis. Note that if $U_{\mathrm{%
R}}=-U_{\mathrm{L}}$ then there is a symmetry $\left\vert \pi _{0}\left(
\mathrm{R}\right) \right\vert =\left\vert \pi _{0}\left( \mathrm{L}\right)
\right\vert _{p_{0}\rightarrow -p_{0}}$ with respect to the change $%
p_{0}\rightarrow -p_{0}$. The solutions $_{\;\zeta }\psi _{n}\left( X\right)
$ and $^{\;\zeta }\psi _{n}\left( X\right) $ are orthonormalized with
respect to the inner product on the $x=\mathrm{const}$ hyperplane, \emph{\ }%
\begin{eqnarray}
&&\left( \ _{\zeta }\psi _{n},\ _{\zeta ^{\prime }}\psi _{n^{\prime
}}\right) _{x}=\zeta \eta _{\mathrm{L}}\delta _{\zeta ,\zeta ^{\prime
}}\delta _{n,n^{\prime }},\ \ \eta _{\mathrm{L}}=\mathrm{sgn\ }\pi
_{0}\left( \mathrm{L}\right) ,  \notag \\
&&\left( \ ^{\zeta }\psi _{n},\ ^{\zeta ^{\prime }}\psi _{n^{\prime
}}\right) _{x}=\zeta \eta _{\mathrm{R}}\delta _{\zeta ,\zeta ^{\prime
}}\delta _{n,n^{\prime }},\ \ \eta _{\mathrm{R}}=\mathrm{sgn\ }\pi
_{0}\left( \mathrm{R}\right) \ ;  \notag \\
&&\left( \psi ,\psi ^{\prime }\right) _{x}=\int \psi ^{\dag }\left( X\right)
\gamma ^{0}\gamma ^{1}\psi ^{\prime }\left( X\right) dtd\mathbf{r}_{\bot }\ ,
\label{c3}
\end{eqnarray}%
if
\begin{align}
& ^{\zeta }\mathcal{N}=\ ^{\zeta }CY,\ _{\zeta }\mathcal{N}=\ _{\zeta }CY,\
Y=(V_{\perp }T)^{-1/2},  \notag \\
& ^{\zeta }C=\left[ 2\left\vert p^{\mathrm{R}}\right\vert \left\vert \pi
_{0}(\mathrm{R})-\chi p^{\mathrm{R}}\right\vert \right] ^{-1/2},\ _{\zeta }C=%
\left[ 2\left\vert p^{\mathrm{L}}\right\vert \left\vert \pi _{0}(\mathrm{L}%
)-\chi p^{\mathrm{L}}\right\vert \right] ^{-1/2}.  \label{L4}
\end{align}%
We consider our theory in a large space-time box that has a spatial volume $%
V_{\bot }=\prod\limits_{j=2}^{D}K_{j}$\ and the time dimension $T$, where
all $K_{j}$\ and $T$\ are macroscopically large. It is supposed that all the
solutions $\psi \left( X\right) $\ are periodic under transitions from one
box to another. Then the integration in (\ref{c3}) over the transverse
coordinates is fulfilled from $-K_{j}/2$\ to $+K_{j}/2$, and over the time $t
$\ from $-T/2$\ to $+T/2$. Under these suppositions, the inner product (\ref%
{c3}) does not depend on $x$.

It is assumed that each pair of solutions $_{\zeta }\psi _{n}\left( X\right)
$ and $^{\zeta }\psi _{n}\left( X\right) $ with any $n\in \Omega _{1}\cup
\Omega _{3}\cup \Omega _{5}$ is complete in the space of solutions with the
corresponding $n$. Mutual decompositions of such solutions have the form (%
\ref{rel1}). The decomposition coefficients $g$ have the following origin:
\begin{equation}
\left( \ _{\zeta }\psi _{n},\ ^{\zeta ^{\prime }}\psi _{n^{\prime }}\right)
_{x}=\delta _{n,n^{\prime }}g\left( \ _{\zeta }\left\vert ^{\zeta ^{\prime
}}\right. \right) ,\ \ g\left( \ ^{\zeta ^{\prime }}\left\vert _{\zeta
}\right. \right) =g\left( \ _{\zeta }\left\vert ^{\zeta ^{\prime }}\right.
\right) ^{\ast },\ \ n\in \Omega _{1}\cup \Omega _{3}\cup \Omega _{5}\ .
\label{c12}
\end{equation}%
They satisfy the unitary relations%
\begin{eqnarray}
&&\left\vert g\left( _{-}\left\vert ^{+}\right. \right) \right\vert
^{2}=\left\vert g\left( _{+}\left\vert ^{-}\right. \right) \right\vert
^{2},\;\left\vert g\left( _{+}\left\vert ^{+}\right. \right) \right\vert
^{2}=\left\vert g\left( _{-}\left\vert ^{-}\right. \right) \right\vert ^{2},
\notag \\
&&\frac{g\left( _{+}\left\vert ^{-}\right. \right) }{g\left( _{-}\left\vert
^{-}\right. \right) }=\frac{g\left( ^{+}\left\vert _{-}\right. \right) }{%
g\left( ^{+}\left\vert _{+}\right. \right) },\ \left\vert g\left(
_{+}\left\vert ^{-}\right. \right) \right\vert ^{2}-\left\vert g\left(
_{+}\left\vert ^{+}\right. \right) \right\vert ^{2}=-\eta _{\mathrm{L}}\eta
_{\mathrm{R}}.  \label{UR}
\end{eqnarray}

In the framework of a field theory an observable $F$\ can be realized as an
inner product of the type (\ref{t4}) of localizable wave packets $\psi (X)$\
and $\hat{F}\psi ^{\prime }(X)$,%
\begin{equation}
F\left( \psi ,\psi ^{\prime }\right) =\left( \psi ,\hat{F}\psi ^{\prime
}\right) ,  \label{A1}
\end{equation}%
where $\hat{F}$\ is a differential operator and $\psi (X)$\ and $\psi
^{\prime }(X)$\ are solutions of the Dirac equation. Assuming that an
observable $F\left( \psi ,\psi ^{\prime }\right) $\ is time-independent
during the time $T$\ one can represent this observable in the following form
of an average value over the period $T$:{\large \ }%
\begin{equation}
\left\langle F\right\rangle =\frac{1}{T}\int_{-T/2}^{+T/2}F\left( \psi ,\psi
^{\prime }\right) dt.  \label{A2}
\end{equation}%
In general the wave packets $\psi (X)$\ and $\psi ^{\prime }(X)$\ can be
decomposed into plane waves $\psi _{n}(X)$\ and $\psi _{n}^{\prime }(X)$\
with given $n$,
\begin{equation}
\psi (X)=\sum_{n}\alpha _{n}\psi _{n}(X),\;\psi ^{\prime }(X)=\sum_{n}\alpha
_{n}^{\prime }\psi _{n}^{\prime }(X),  \label{A3}
\end{equation}%
where $\psi _{n}(X)$\ and $\psi _{n}^{\prime }(X)$\ are superpositions of
the solutions$_{\;\zeta }\psi _{n}\left( X\right) $\ and $^{\;\zeta }\psi
_{n}\left( X\right) $. Taking into account the orthogonality relation (\ref%
{c3}) one finds that the decomposition of $\left\langle F\right\rangle $\
into plane waves with given $n$\ does not contain interference terms,

\begin{equation}
\left\langle F\right\rangle =\sum_{n}F\left( \alpha _{n}\psi _{n},\alpha
_{n}^{\prime }\psi _{n}^{\prime }\right) .  \label{A4}
\end{equation}

\end{document}